\documentclass[prb,twocolumn,aps,showkeys,superscriptaddress]{revtex4-1}
\usepackage{amsmath}
\usepackage{amsfonts}
\usepackage{amssymb}
\usepackage{graphicx}
\usepackage{epstopdf}
\usepackage{float} 
\usepackage[caption=false]{subfig}
\usepackage[usenames,dvipsnames,svgnames,table]{xcolor} 
\usepackage[normalem]{ulem} 
\usepackage[makeroom]{cancel} 
\usepackage{tikz} 

\DeclareMathOperator{\Li}{Li}
\DeclareMathOperator{\Tr}{Tr}

\begin{document}

\title{Phase transitions and pattern formation in ensembles of phase-amplitude solitons in quasi-one-dimensional electronic systems.}
\author{P. Karpov}
\email{karpov.petr@gmail.com}
\affiliation{Max Planck Institute for the Physics of Complex Systems, N{\"o}thnitzer Stra{\ss}e 38, Dresden 01187, Germany}
\affiliation{National University of Science and Technology ``MISiS'', Moscow, Russia}
\author{S. Brazovskii}
\affiliation{CNRS UMR 8626 LPTMS, University of Paris-Sud, University of Paris-Saclay, Orsay, France}
\affiliation{National University of Science and Technology ``MISiS'', Moscow, Russia}
\affiliation{Jozef Stefan Institute, Jamova 39, SI-1000 Ljubljana, Slovenia}


\begin{abstract}
Most common types of symmetry breaking in quasi-one-dimensional electronic systems possess a combined manifold of states degenerate with respect to both the phase $\theta$ and the amplitude $A$ sign of the order parameter $A\exp(i\theta)$. These degrees of freedom can be controlled or accessed independently via either the spin polarization or the charge densities. To understand statistical properties and the phase diagram in the course of cooling under the controlled parameters, we present here an analytical treatment supported by Monte Carlo simulations for a generic coarse-grained two-fields model  of XY-Ising type. The degeneracies give rise to two coexisting types of topologically nontrivial configurations: phase vortices and amplitude kinks -- the solitons. In 2D, 3D states with long-range (or BKT type) orders, the topological confinement sets in at a temperature $T=T_1$ which binds together the kinks and unusual half-integer vortices. At a lower $T=T_2$, the solitons start to aggregate into walls formed as rods of amplitude kinks which are ultimately terminated by half-integer vortices.
With lowering $T$, the walls multiply passing sequentially across the sample.

The presented results indicate a possible physical realization of a peculiar system of half-integer vortices with rods of amplitude kinks connecting their cores. Its experimental realization becomes feasible in view of recent successes in real space observations and even manipulations of domain walls in correlated electronic systems.
\end{abstract}

\date{\today}
\maketitle

\section{Introduction}

Symmetry broken electronic states give rise to topological defects: from extended objects like vortices or domain walls as solitonic lattices to microscopic solitons as anomalous quasi-particles and instantons in their dynamics.
Role of solitons in electronic properties was appreciated in theories since mid 70's (see reviews \cite{BK:84,YuLu}, and also \cite{braz-moriond,JSSS08,SB-Landau100:2009} for a more recent development). The solitons were firstly accessed in experiments on conducting polymers of early 80's \cite{Heeger-RMP}. New motivations came in early 2000's from discoveries of the ferroelectric charge ordering (see reviews \cite{brazov-springer, Monceau-springer,Monceau-advances}) in organic conductors \cite{lebed-springer}, from new accesses via nano-scale tunneling experiments \cite{latyshev-prl:05,latyshev-prl:06} in materials with Charge Density Waves (CDW), from optics of new conducting polymers \cite{vardeni}.

Most clearly the existence of solitons was confirmed by recent observations in CDWs where the individual solitons have been visually captured by STM experiments \cite{Brun-soliton,Kim:2012}. There were amplitude solitons (through which the order parameter changes the sign) corresponding to the long-sought special quasi-particles -- the spinons or the holons.
New opportunities are coming from the latest facilities of manipulating cooperative electronic states by short optical pulses and by a local STM injection. On this way, the networks of domain walls -- the line aggregates of solitons -- have been created in a layered CDW compound $\mathrm{Ta S_2}$ \cite{Stojchevska:2014,Yoshida:2014,Vaskivskyi:2016,Ma:2016,Cho:2016,Svetin:2016,Gerasimenko:2017,Karpov:2018}.

A major puzzle, as well as the inspiration, coming from the experimental observations, is that the solitons were observed \cite{Brun-soliton} within the low temperature ($T<T_{c}$) phase with the long-range 3D order. The hidden obstacle is the effect of the confinement appearing in higher dimensions $D>1$ \cite{JSSS08,braz-moriond}. Commuting between degenerate minima on only one chain would lead to a loss of the interchain ordering energy proportional to $|x|$ -- the length along the chain till the next soliton. In case of discrete symmetries (when solitons are the amplitude kinks as in the $Z_2$ case of the dimerization) the solitons are bound in topologically trivial pairs with an option for a subsequent phase transition to form cross-sample domain walls \cite{Bohr:1983,Teber:2001,Teber:2002,Karpov:2016}. But for a continuous symmetry (the phase degeneracy in superconductors (SC) and incommensurate charge density waves (ICDW) or the directional degeneracy in spin-isotropic antiferromagnets (AFM)), both phase and spin degeneracy in spin density waves (SDW)) the gapless mode can cure the interruption from the amplitude kink.

As we shall explain below, continuously broken symmetries  allow for individual solitons entering the low-temperature phases with long- range ordered states. The solitons adapt by forming topologically bound complexes with half-integer vortices of gapless modes: $\pi$-rotons \cite{braz-moriond,JSSS08,SB-Landau100:2009}. For cases of repulsion and attraction correspondingly, that results in the spin- or charge- roton configurations with charge- or spin- amplitude kinks localized in the core.
Beyond the quasi-1D electronic systems, the interference of amplitude and phase (or angle) topological defects has  been noticed already for superfluid phases of surface layers of $\mathrm{He_3}$ [\onlinecite{Korshunov:1985}] which provoked the studies of different generalizations of the XY model \cite{Lee:1985,Korshunov:1986,Shi:2011,Serna:2017}.

In this paper, we construct and study the coarse-grained model of XY-Ising type, which describes
local and the long-range ordering under the monitored mean concentration of solitons.
The paper is organized as follows. In Sec. \ref{Solitons in quasi-1D systems} we set the playground: we introduce the general concept of solitons and, in particular, of solitons in quasi-1D electronic cooperative.
In Sec. \ref{sec_analytical} we present the discretized version of the model and perform its qualitative analysis in various regimes. Section \ref{Sec Numerical simulations} describes the results of numerical simulations for 2D and 3D systems. Finally, in Sections \ref{Sec Discussion} and \ref{Sec Conclusions} we present the Discussion and draw the Conclusions.

\section{Solitons in quasi-1D systems.}
\label{Solitons in quasi-1D systems}

At the 1D level, for systems with a complex order parameter $\mathcal{O}=A\exp(i\varphi)$, the amplitude soliton $A(x=-\infty)\Rightarrow-A(x=+\infty)$ performs the sign change  of the order parameter $\mathcal{O}$ at an arbitrary $\varphi=const$. At higher space and/or space-time dimensions, the requirement appears that the order parameter is mapped onto itself. Even being favorable in energy in comparison with the band electron, the soliton cannot be created dynamically already in $D=1$ space dimension, and is prohibited to exist even stationary at $D>1$. The resolution is to invoke the combined symmetry: the amplitude kink $A\Rightarrow-A$ coupled with the half-integer $\varphi\Rightarrow\varphi\pm\pi$ vortex of the phase rotation which compensates for the amplitude sign change. The resulting kink-roton complex, illustrated in Figs. \ref{fig:holon}, \ref{fig:string} allows for several interpretations in applications to ICDW or SC: 2D view is a pair of $\pi$-vortices sharing the common core bearing one unpaired spin which stabilizes the state; 3D view is a ring of a half-flux vortex line, its center confines the spin; at any $D>1$ this is a nucleus of the melted Fulde-Ferrel-Larkin-Ovchinnikov phase in the spin-polarized superconductors \cite{Fulde:1964,Larkin:1964}, or its CDW analogue \cite{Buzdin,SB-Landau100:2009}.

\subsection{Solitons in incommensurate CDWs}

\subsubsection{Amplitude solitons in 1D.}

Here we quote some basic facts on microscopic origination of solitons in ICDW, see \cite{BK:84,braz-moriond,JSSS08,SB-Landau100:2009}, based upon the Peierls-Froelich model with generalization to BCS case for filamentary superconductors \cite{Buzdin,Machida}. The ICDW is a crystal $A\cos(2K_{f}x+\varphi)$ of electron-hole pairs which order parameter is
$\mathcal{O}_{ICDW}=A\exp(i\varphi)$.
The singlet pair can be broken into spin $1/2$ components, but  that will not be an expectedly liberated electron-hole pair at the gap rims $\pm\Delta_{0}$. Rather, there will be spin-carrying ``amplitude solitons'' (AS) -- nodes of the order parameter distributed over the length $\xi_{0}=\hbar v_F/\Delta_{0}$, see \cite{BK:84,braz-moriond,JSSS08,SB-Landau100:2009}. The total energy of the soliton is $\approx2\Delta_{0}/3<\Delta_{0}$ making it energetically favorable with respect to the energy $\Delta_0$ of the undressed unpaired electron which now is trapped at the midgap state associated with the amplitude soliton. This electron brings the spin $1/2$, but the charge of the AS is compensated to zero by the local dilatation in the filled manifold of singlet vacuum states \cite{BK:84}. That makes the AS be a CDW realization of the ``spinon'', with a direct generalization to a superconductor \cite{kwon:02}.

\subsubsection{An unpaired spin in the ICDW or SC environment at $D>1$.}

As a nontrivial topological object (the order parameter $O_{cdw}(x)=A(x)\exp(i\varphi(x))$ does not map onto itself), the pure AS (with $\varphi(x)=const$) is prohibited in $D>1$ environment. Nevertheless, the AS becomes allowed if it acquires phase tails with the total increment $\delta\varphi=\pm\pi$. The
length of these tails $\xi_{\varphi}\gg\xi_{0}=\hbar v_F/\Delta_0$ is determined by the weak interchain coupling. As in 1D, the sign of $A(x)$ changes within the scale $\xi_{0}$ but further on, at the scale $\xi_{\varphi}$, the factor $\exp(i\varphi)$ also changes the sign, thus leaving the product in $O_{cdw}$ to be invariant. As a result, the 3D allowed particle is formed with the AS core $\xi_{0}$ carrying the spin, and the two $\pi/2$-twisting wings stretched over $\xi_{\varphi}$, each carrying the charge $e/2$. This picture can be directly reformulated for quasi-1D superconductors by redefining the meaning of the phase. The phase tails form now the elementary $\pi$-junction \cite{braz-moriond,kwon:02}.

\subsection{Solitons within the bosonisation language}

\subsubsection{Solitons in spin-singlet systems.}

Recall a universal microscopic insight to excitations in 1D electronic systems, first in application to the spin-gap cases -- SC or CDW. The starting single chain level is well described by the bosonization language, see \cite{Giamarchi:book,Tsvelik:book}.
The Hamiltonian
\begin{align}
H_{1D}=&(\hbar/4\pi)[v (\partial_x\theta)^{2}+(\partial_t\theta)^{2}/v] - \nonumber\\
&-V\cos(2\theta) +(\hbar/4\pi\gamma)[v(\partial_x\varphi)^{2}+(\partial_t\varphi)^{2}/v]
\label{eq_H1D}
\end{align}
is written in terms of phases $\theta$ and $\varphi$ for spin and charge degrees of freedom correspondingly. The energy $\sim V$ comes from the backward exchange scattering $V\sim g_{1}$ of electrons. The pair-breaking excitation - the $s=1/2$ spinon, is the soliton connecting the degenerate minima of $H_{1D}$:
$\theta\Rightarrow\theta \pm\pi$. The singlet order parameter, for either SC or CDW (with the inversion of the parameter $\gamma$ depending on a definition of the charge phase $\varphi$) is
$\mathcal{O}_{SC,CDW}\sim\cos\theta\exp(i\varphi)$.
Its amplitude $A=\cos\theta$ changes the sign across the allowed $\pi$ soliton in ${\theta}$, hence the spinon is an alternative description of the same amplitude soliton which appears in BCS-Peierls type models.

For the ensemble of chains, labeled below by indices $\alpha,\beta$, the relevant interchain coupling reads
\begin{align}
&\int dx\sum_{\alpha,\beta}J_{\alpha,\beta}\mathcal{O}_\alpha\mathcal{O}^*_\beta = \nonumber\\
& =\int dx\sum_{\alpha,\beta}J_{\alpha,\beta}\cos\theta_\alpha \cos\theta_\beta
\exp(i(\varphi_\alpha-\varphi_\beta))
\label{CDW-perp}
\end{align}

\subsubsection{Solitons in the Mott insulator: SDW route.}

Consider the quasi-1D system with repulsion at a nearly half-filled band, which is the SDW route to a general doped antiferromagnetic  Mott-Hubbard insulator \cite{SB-Landau100:2009}.
The single-chain state has no magnetic long-range order, gapeless spin degrees of freedom are described by the same phase $\theta$ as in (\ref{eq_H1D}), (\ref{CDW-perp}), charge degrees of freedom are described by the chiral phase $\varphi(x,t)$ corresponding to the CDW version of (\ref{eq_H1D}), (\ref{CDW-perp}). The 1D bosonized Hamiltonian \cite{Giamarchi:book,Tsvelik:book} becomes $H_{1D}=H_c+H_s$
\begin{align}
H_c(\varphi) &=(\hbar/4\pi\gamma)[v_c(\partial_{x}\varphi)^{2}+
(\partial_{t}\varphi)^{2}/v_c]-U\cos(2\varphi), \nonumber\\
H_s(\theta) &= (\hbar/4\pi)[v_s(\partial_{x}\theta)^{2}+(\partial_{t}\theta)^{2}/v_s]
\label{H1D-sdw}
\end{align}
Here $v_c$ and $v_s$ are the charge and spin phase velocities, $U$ is the Umklapp scattering amplitude  $U\sim g_3$ which is responsible for the charge gap formation \cite{DL,Luther+Emery}. The constant $\gamma$ (commonly called the Luttinger liquid parameter -- $K_{\rho}$,  dependent on interactions) determines several regimes: $\gamma<1$ at any repulsion, then $U$ is preserved if it is already build-in - this is the generic Mott state case for a half-filled band; $\gamma<1/2$ at a stronger repulsion, then $U$ is spontaneously generated even away from the bare half-filling - this is the Charge Ordering case \cite{brazov-springer}.

In 1D, the excitations are the \textit{holon/doublon} pairs seen as the $\mp\pi$ solitons in $\varphi$, and the spin sound in $\theta$, which are decoupled.
In a quasi-1D system, multiple forms of the interchain coupling appear among which the relevant one is the coupling via the SDW order parameter: the staggered magnetization
$\mathcal{O}_{sdw} \propto S_x+iS_y \propto \cos\varphi\exp(i\theta)$.

For the ensemble of chains, analogously to (\ref{CDW-perp}), the relevant interchain coupling reads
\begin{equation}
\label{SDW-perp}
\int dx\sum_{\alpha,\beta}J_{\alpha,\beta}\cos\varphi_\alpha \cos\varphi_\beta
\exp(i(\theta_\alpha-\theta_\beta))
\end{equation}
We see that the $\pi$-soliton in $\varphi$ becomes the amplitude kink ($\cos\varphi\rightarrow-\cos\varphi$), and to survive in $D>1$ it should enforce the $\pi$ rotation in $\theta$, then the sign changes in both of the two factors composing $\mathcal{O}_{sdw}$ compensate and the configuration becomes allowed.

\begin{figure}[tbh]
\subfloat[]{\includegraphics[height=2.7cm]{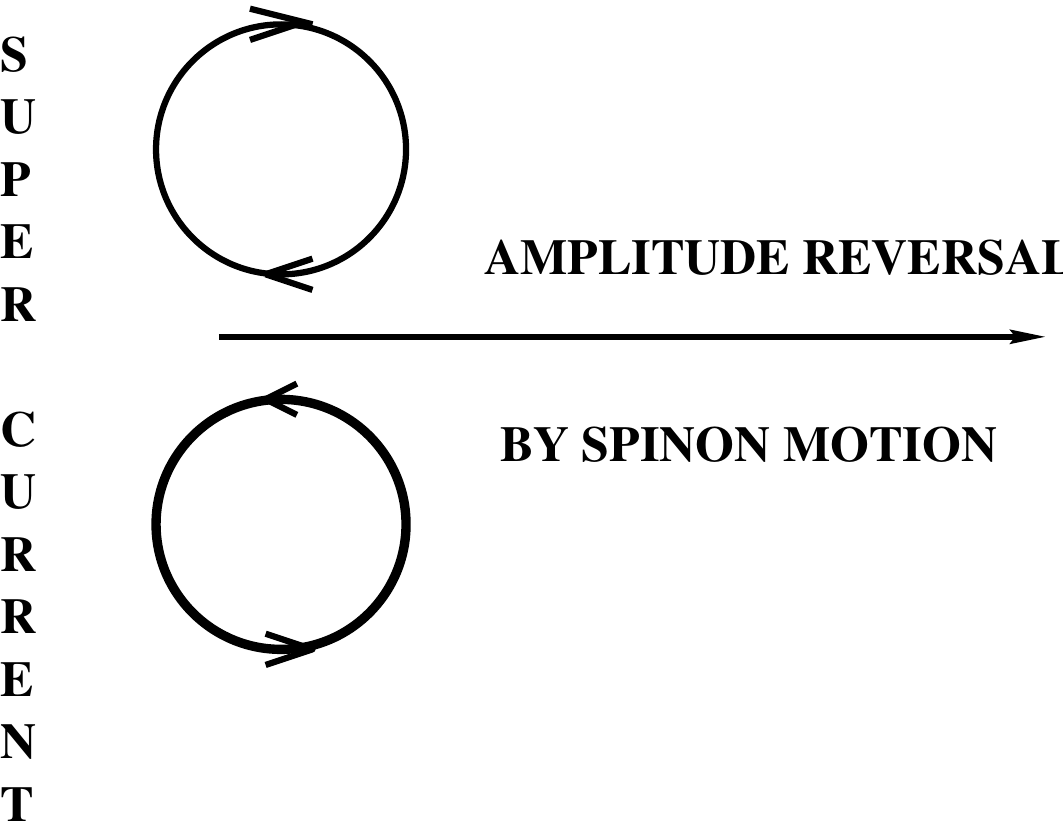}}
\quad
\subfloat[]{\includegraphics[height=2.7cm]{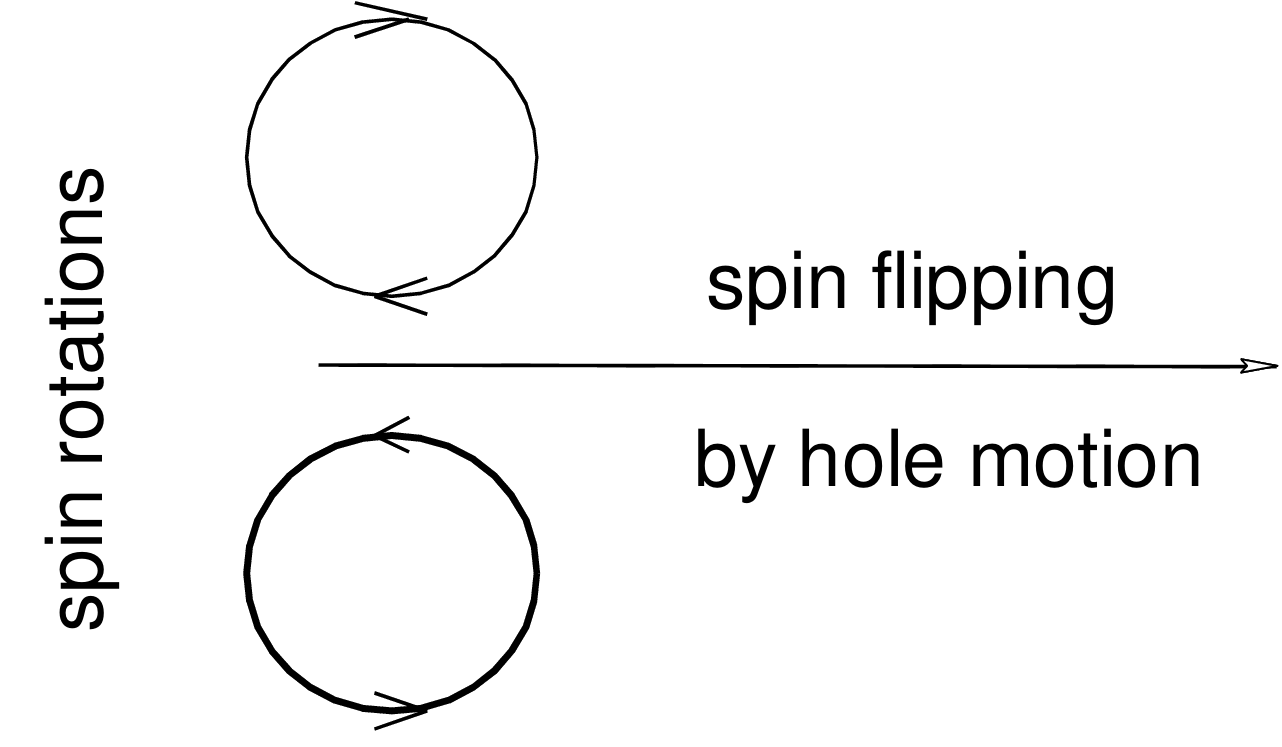}}
\caption{Motion of the kink-roton complexes. (a) For the ICDW or the superconductor, the amplitude kink plays the role of the spinon. For the ICDW the curls are displacements contours for the half-integer dislocation pair. For the superconductor, the curls are lines of electric currents circulating through the normal core carrying the unpaired spin.
(b) For the SDW or the AFM, the string of the amplitude reversal
of the order parameter created by the holon is cured by the semi-vortex pair of the staggered magnetization circulation.  }
\label{fig:holon}
\end{figure}

This construction can be generalized beyond quasi-1D systems by considering a vortex configuration bound to an unpaired electron. Extending to AFMs like the pristine $\mathrm{CuO_2}$ planes, the SDW becomes a staggered magnetization; the soliton becomes a hole, which motion leaves the string of reversed AFM sublattices \cite{Bulaevskii:1968,Brinkman:1970}; the $\pi$-wings become the magnetic semi-vortices. The resulting configuration is a half-integer vortex ring of the staggered magnetization (a semi-roton) with the holon confined in its center (see  Fig. \ref{fig:holon}b  -- in 2D when the vortex ring is reduced to the pair of vortices.)
Such a combined semi-vortex -- Fig. \ref{fig:holon} might be significant also in sliding incommensurate SDWs which possesses an even richer order parameter \cite{NK+SB:00}.

\subsection{Inverse rout: from stripes to solitons and fractional vortices.}

\begin{figure}[tbh]
 \includegraphics[width=0.4\textwidth]{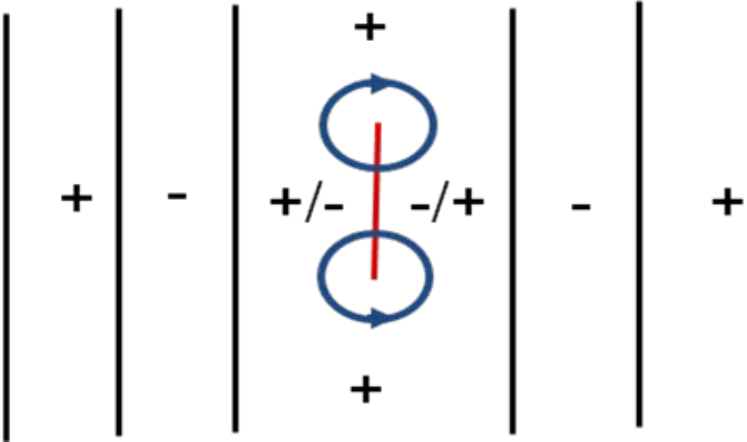}\\
 \caption{Kink-roton complexes as nucleuses of  melting of the stripes lattice. The defect is embedded into the regular stripe structure (black lines for the amplitude zeros); $+/-$ are the alternating signs of the order parameter $\mathcal{O}$ amplitude. Termination points of a finite segment $L_{\perp}$ (red color) of the $\mathcal{O}=0$ line must be encircled by semi-vortices of the phase rotation (blue circles) to resolve the signs mismatch. The minimal segment corresponds to the elementary kink carrying spin 1/2.}
 \label{fig:string}
\end{figure}

At low temperature and at large concentration, the amplitude solitons tend to organize themselves macroscopically, as a periodic array of stripes crossing the whole sample in the transverse direction. The arrays of spin solitons have been identified experimentally e.g. in spin-Peierls systems \cite{Horvatic:1999}. No confinement energy is lost in this perfect state since the order parameter alternates coherently across all chains.  But when the solitonic lattice melts,
the defects appear such as termination points of lines of the amplitude nodes. To be allowed, each such a point should be complemented by the pair of $\pi$-vortices, Fig. \ref{fig:string}.

\section{Formulation of the effective model and its qualitative analysis}
\label{sec_analytical}

A discrete version of Hamiltonians (\ref{eq_H1D}), (\ref{CDW-perp}) or (\ref{H1D-sdw}), (\ref{SDW-perp}) can be written as \cite{Bohr:1983}
\begin{align}
H =& - A_{||} \sum_r \cos(\theta_{r}-\theta_{r+\hat{x}}) - \nonumber \\
&-A_{\perp} \sum_{r} s_{r} s_{r+\hat{y}} \cos(\theta_{r}-\theta_{r+\hat{y}}) - J_{||} \sum_{r} s_{r} s_{r+\hat{x}}
\label{HamiltonianIsing}
\end{align}
Here the Ising variable $s=\pm1$ corresponds to the normalized real amplitude of the order parameter and the angle $\theta$ corresponds to the phase of the complex order parameter; $r$ runs over sites of a square lattice. The last term fixes the chemical potential $\mu_s$ of amplitude solitons (and therefore their average number in the system, see \cite{Karpov:2016} for the details).
For the last constraint to be held, the constant $J_{||}=J_{||}(\nu, T) = (E_s-\mu_s(\nu,T))/2$ has to be found self-consistently by maintaining the average concentration (per site) of solitons $\nu = -\partial \Omega/\partial \mu_s = T \partial \ln Z/\partial \mu_s$.
The discretization length $a$  for integrals (\ref{CDW-perp}) or (\ref{SDW-perp}) is chosen to be equal to the minimal in-chain distance between solitons, which is determined by the short-range repulsion of solitons at a given chain.

The Hamiltonian (\ref{HamiltonianIsing}) is invariant under a symmetry (gauge) transformation of simultaneous flipping of all Ising and all XY spins within any given chain:
\begin{align}
    \begin{cases}
    s_{x,y} \rightarrow -s_{x,y}   \qquad  & \mbox{for all $s, \theta$} \\
    \theta \rightarrow \theta+\pi  \qquad & \quad   \mbox{at a given chain $y$}
    \end{cases}
\label{eq:gauge_transformation}
\end{align}
which implies vanishing of macroscopical averages of Ising and XY fields separately: $\langle s_r\rangle=0$, $\langle e^{i\theta_r} \rangle=0$. Still a combined  variable $\Delta_r= s_r  e^{i\theta_r} $ can have nontrivial average and correlation functions.

Now we give a brief ``dictionary''  for different terms we use throughout the paper from the point of view of the discretized Hamiltonian (\ref{HamiltonianIsing}) the Ising variables $s_r$.

An \textit{amplitude soliton} (of just a \textit{soliton} for brevity) is an interruption of the Ising order along the chains' ($x$) direction, i.e. the soliton is present when $s_r s_{r+\hat{x}}=-1$.

A \textit{bisoliton} consists of two solitons at the minimal in-chain distance $a$ between them; in the language of the Ising variables $s_r$ it is a reversed spin in the otherwise ordered part of the chain.

A \textit{solitonic rod} (or a \textit{solitonic line}) is transverse to the chains aggregation of solitons, when solitons at several neighboring chains have the same $x$-coordinate. Generalized to the 3D case such a structure would be called a \textit{solitonic disk}. For the transverse aggregate of bisolitons instead of solitons, we accordingly use terms \textit{bisolitonic rod} (in 2D) or \textit{bisolitonic disk} in 3D.

A \textit{solitonic wall} can appear when a solitonic line grows and crosses the whole transverse section of the sample. Likewise, a \textit{bisolitonic wall} appears, when a bisolitonic rod grows across the sample.

\subsection{Zero chemical potential of amplitude solitons}
\label{Summing-out-Ising-variables}

We start studying the model (\ref{HamiltonianIsing}) from the simple, but important case $J_{||}=0$. That corresponds to the zero chemical potential of solitons (with respect to their activation energy), maintained by a
solitonic reservoir, with which they can exchange the solitons with no energy cost. For an isolated system, the role of such a reservoir is played by the solitonic domain walls \cite{Bohr:1983,Teber:2001,Karpov:2016}.

For $J_{||}=0$, the model becomes an anisotropic version of the XY-Ising model \cite{Yosefin:1985}:
\begin{align}
H_0 (A_{||}, A_{\perp}) =& - A_{||} \sum_r \cos(\theta_{r}-\theta_{r+\hat{x}}) - \nonumber \\
&-A_{\perp} \sum_{r} s_{r} s_{r+\hat{y}} \cos(\theta_{r}-\theta_{r+\hat{y}})
\label{HamiltonianIsingJ=0}
\end{align}
with the XY interaction only along the chains' $x$-direction and XY-Ising interaction only along the inter-chain $y$-direction.

Using the identity $s_r \cos(\theta)=\cos(\theta+(1-s_r)\pi/2)$ we can show that the model is self-dual with respect to transformation of the angle variable $\chi_r = \theta_r+(1-s_r)\pi/2$. Indeed, the Hamiltonian (\ref{HamiltonianIsingJ=0}) transforms to
$$
H_0=-A_{||} \sum_r s_{r} s_{r+\hat{x}} \cos(\chi_{r}-\chi_{r+\hat{x}}) -A_{\perp} \sum_{r}  \cos(\chi_{r}-\chi_{r+\hat{y}}),
$$
which is equivalent to (\ref{HamiltonianIsingJ=0}) under the substitutions $\theta\leftrightarrow\chi$, $A_{||} \leftrightarrow A_{\perp}$, $ x \leftrightarrow y$. This means that all thermodynamic functions have to be symmetric with respect to the interchange $A_{||} \leftrightarrow A_{\perp}$. Also, the correlation functions have to be symmetric with respect to the simultaneous change  $A_{||} \leftrightarrow A_{\perp}$, $x \leftrightarrow y$, for example
$$
\Tr \{s_r s_{r+\hat{x}} e^{-\beta H(A_{||}, A_{\perp})} \}= \Tr\{ s_r s_{r+\hat{y}} e^{-\beta H(A_{\perp}, A_{||}) }\}.
$$
Particularly, in the limit $A_{\perp}=A_{||}$ we have $\langle s_r s_{r+\hat{x}} \rangle = \langle s_r s_{r+\hat{y}} \rangle$ .

The summation over the Ising variables in the partition function generated by the Hamiltonian (\ref{HamiltonianIsingJ=0}) can be performed exactly.
For a fixed configuration of the XY field $\theta_r$ in (\ref{HamiltonianIsingJ=0}), the Hamiltonian describes a set of non-interacting transverse Ising chains with random interactions between the neighboring spins $s_r$ across the chain.

Let $s_r s_{r+\hat{y}} = \tilde{s}_r = \pm 1$, then the Hamiltonian (\ref{HamiltonianIsingJ=0}) becomes $H_0 = - A_{||} \sum_r \cos(\theta_{r}-\theta_{r+\hat{x}}) - A_{\perp} \sum_{r} \tilde{s}_{r} \cos(\theta_{r}-\theta_{r+\hat{y}})$, where $\tilde{s}_r$ are independent Ising variables.
Summing out $\tilde{s}_r$ we arrive at the effective Hamiltonian for the modified XY-field
\begin{align}
\tilde{H}_0 =& - A_{||} \sum_r \cos(\theta_{r}-\theta_{r+\hat{x}}) \nonumber\\
&- \frac{1}{\beta} \sum_r \ln \cosh (\beta A_{\perp} \cos(\theta_{r}-\theta_{r+\hat{y}}))
\label{H-Korsh}
\end{align}
The interaction is fundamentally anisotropic: it is $2\pi$-periodic in the chains' ($x$) direction, and only $\pi$-periodic
in the transverse ($y$) direction. That differs from the isotropic Korshunov model \cite{Korshunov:1985,Lee:1985,Korshunov:1986}, where both $\pi$ and $2\pi$ interactions coexist in all directions. In the low-temperature limit $\beta A_{\perp}\gg1$, the Hamiltonian (\ref{H-Korsh}) becomes
\begin{align}
\tilde{H}_0 = - A_{||} \sum_r \cos(\theta_{r}-\theta_{r+\hat{x}}) - A_{\perp} \sum_{r} \left| \cos(\theta_{r}-\theta_{r+\hat{y}}) \right|
\label{H-abs}
\end{align}
For slowly varying configurations (apart from the transverse $\pi$-jumps in $\theta_r$),  keeping only the lowest order periodicities in increments of $\theta$, the Hamiltonians (\ref{H-Korsh}), (\ref{H-abs}) can be mapped upon the form
\begin{align}
\tilde{H}_0 = - A_{||} \sum_r \cos(\theta_{r}-\theta_{r+\hat{x}}) - \frac{A_{\perp}}{4} \sum_{r} \cos 2(\theta_{r}-\theta_{r+\hat{y}})
\end{align}

The isotropic version of this model (i.e. when both $\theta$- and $2\theta$-terms are present in both $x$ and $y$ directions simultaneously) has been studied in \cite{Korshunov:1985,Lee:1985,Korshunov:1986}. For $A_{||} \ll A_{\perp}$ it gives rise to half-integer $\pi$-vortices (allowed by the $\sim A_{\perp}$ term) which are bound in pairs by ``strings''  generated by the $2\theta$ term (here, the strings are borders separating $\pi$-mismatches of $\theta$ \cite{Korshunov:1985,Lee:1985,Korshunov:1986}). But for the model we study here, where each type of terms is associated with a particular direction, the situation is different.
Even when $A_{||} \simeq A_{\perp}$, the model allows for $\pi$-vortices with strings attached in $x$-direction (so that $\pi$-mismatches in phase $\theta$ take place only in $y$-direction, for an example see  Fig. \ref{Fig3-HalfV}).
These $\pi$-vortices take the role of the amplitude solitons which are hiddenly present in the excluded Ising variables. The $A_{||}$ term still provides the confinement of $\pi$-vortices into pairs, but only in the $y$-direction.

\subsection{Fixed non-zero chemical potential of solitons}

Now consider the complete Hamiltonian (\ref{HamiltonianIsing}) with $J_{||} \neq 0$,
putting for simplicity $A_{||} = A_{\perp} \equiv A$.
Using the Villain approximation we can transform (\ref{HamiltonianIsing}) to the Coulomb gas representation as described in Appendix A:
\begin{align}
Z &= \sum_{\{m_R\}} \sum_{\nu_R(\{s\})} \nonumber \\
&\exp\left( \pi \beta A \sum_{R\neq R'} (m_R + \nu_R) \ln\frac{|{\mathbf R}-{\mathbf R}'|}{a} (m_{R'} + \nu_{R'}) \right.\nonumber \\
&\left. \phantom{ \ln\frac{{\mathbf R}}{a}} + (\ln y_0)\sum_R (m_R+\nu_R)^2 + \beta J_{||}  \sum_{r} s_r s_{r+\hat{x}} \right).
\label{eq-Z}
\end{align}
Here $a$ is a lattice spacing cutoff, $y_0 = \exp(-\pi^2 \beta A/2)$, summation goes over all integer values of $m_R$ and all configurations of spins $s_r=\pm 1$, the index $r$ corresponds to the original lattice and $R$ corresponds to the dual one; half-integers $\nu_R =  \frac{1}{4}(s_{r+\hat{x}} s_{r+\hat{x}+\hat{y}} - s_r s_{r+\hat{y}})$.
The vortex ``charge'' $m_R+\nu_R$ becomes dependent on the local configuration of spins. Also apart from the long-range ``Coulomb'' energy associated with half-vortices, there is an additional local Ising-term $\sim J_{||}$. This term is important, because as long as $J_{||} \neq 0$ it leads to confinement of the transverse half-vortex pairs, situated at the ends of solitonic rods (see also Sec. \ref{Intermediate beta Jpar and the phase diagram}).

Since the fugacity $y_0$ is small, we can take into account only the leading terms with minimal vorticities which are $m_R=0$ if $\nu_R=0$, $m_R=0,-1$ if $\nu_R=+1/2$, $m_R=0,+1$ if $\nu_R=-1/2$.


\subsection{Qualitative phase diagram}

In this section we reconstruct the phase diagram of the system, based on the qualitative interpretation of partition function (\ref{eq-Z}) corresponding to the Hamiltonian (\ref{HamiltonianIsing}). Again, for simplicity, we put $A_{||} = A_{\perp} = A$.

We start construction of the phase diagram with the limiting cases: the weak coupling limit $\beta J_{||} \rightarrow 0$ (corresponding to the zero chemical potential of solitons, when the Hamiltonian is given by (\ref{H-Korsh})) and the strong coupling limit $\beta J_{||} \rightarrow \infty$ (corresponding to a strongly negative chemical potential of solitons $\mu_s-E_s\rightarrow-\infty$, hence to their low concentration).

\subsubsection{Weak coupling limit for Ising spins $\beta J_{||} \rightarrow 0$}

\begin{figure}[tbh]
	\centering
	\includegraphics[width=0.31\linewidth]{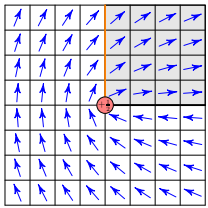}%
	\caption{Single half-vortex. Arrows show XY degree of freedom, white/gray shading shows the Ising degree of freedom.
		The orange vertical thick line shows the rod of the amplitude solitons -- which are energy-costing $J_{||}$ mismatches. The black horizontal thick line shows the ``string'' attached to a half-vortex, i.e. the line of transverse $\pi$-jumps in the phase $\theta$, which do not cost any additional energy.}
	\label{Fig3-HalfV}
\end{figure}

Consider, first, the weak coupling limit for the Ising spins, $\beta J_{||} \rightarrow 0$, which corresponds to the case when the chemical potential of amplitude solitons is trapped at zero (this happens e.g. below the temperature of solitons' walls formation in 3D or in the limit $T\rightarrow0$ in 2D, see Secs. \ref{Subsec_Estimations_for_T1_T2}, G below).
This limit was already discussed in Sec. \ref{Summing-out-Ising-variables}.
Integer vortices disintegrate into pairs of half-integer ones, which lowers the system energy.
A string of XY phase mismatch is attached to every half-integer vortex (Fig. \ref{Fig3-HalfV}).
We note that unlike the model of Korshunov \cite{Korshunov:1985,Lee:1985,Korshunov:1986}, where the strings did cost some energy and thus were uniquely physically defined, in our case,  only the solitonic rod is uniquely defined, while for the string only its $y$-coordinate is fixed.
Still there is an ambiguity from which side, from the left or from the right, the string is attached (for example, applying the gauge transformation (\ref{eq:gauge_transformation}) to all the chains in the upper half-plane in configuration in Fig. \ref{Fig3-HalfV} we will get the string attached to the half-vortex from the left side).
This differs from conventional XY-vortices, where the direction of the ray of $2\pi$-jump of the phase can be chosen completely arbitrarily.
Since the string energy density is zero, then the string does not affect the statistics of the associated vortices. Then, analogously to BKT-case, we expect that at low temperatures all half-vortices are bound into neutral pairs and the transition proceeds via their dissociation. In comparison to the usual BKT-transition, this one has the temperature approximately 4 times lower $T^{half}_{BKT} = T_{BKT}/4$ and the 4 times larger jump of helicity modulus \cite{Korshunov:1986-FFXY}.

\subsubsection{Strong coupling limit for Ising spins  $\beta J_{||} \rightarrow \infty$}
\label{subsec_strong_coupling}

Now consider the opposite,  strong coupling limit for Ising spins $\beta J_{||} \rightarrow \infty$; in terms of the original solitons this means that their chemical potential goes to minus infinity i.e. their concentration vanishes. If all Ising variables were equal (showing the perfect Ising long-range order, i.e. the total absence of amplitude solitons), even then XY spins would show only the algebraic order $\langle e^{i(\theta_r - \theta_{r'})} \rangle \sim |{\mathbf r} - {\mathbf r'}|^{-1/2\pi K}$ at low temperatures.

\begin{figure}[tbh]
\centering
  \includegraphics[width=0.7\linewidth]{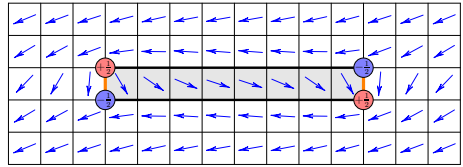}%
\caption{Half-integer vortex complex which breaks Ising order at any $T\neq0$.
In terms of solitons (orange lines) this means that even initially bound soliton-antisoliton pairs will deconfine, lowering the free energy of the system.}
\label{Fig-BreakingIsingOrder}
\end{figure}

But taking into account the gauge symmetry (\ref{eq:gauge_transformation}) it follows that $\langle S_r S_{r'}\rangle = 0$ and $\langle e^{i(\theta_r - \theta_{r'})} \rangle = 0$ for any $r$ and $r'$ lying at different chains, even  at $T=0$, i.e. for $\beta J_{||} \rightarrow \infty$. The natural question then arises: is there Ising order at one given chain at $T\neq0$?
The answer is no: the combined topological defect of a unit height and an arbitrary length (Fig. \ref{Fig-BreakingIsingOrder}) has a finite energy $E$. So applying the known argument for 1D systems \cite{Landau-argument,Landau5} we see that such defects break the in-chain Ising long-range order.

Nevertheless, the long-range correlations can be present for the physically significant combined order parameter $\Delta_r = S_r e^{i\theta_r}$.
That can already be seen from the fact that for $J_{||}\rightarrow\infty$ all the half-vortices are suppressed, and we have the usual Berezinskii-Kosterlitz-Thouless (BKT) transition at the temperature $T_{BKT} \approx 0.9 A$
[\onlinecite{Shugard:1978}, \onlinecite{Olsson:1995}], with $2/\pi$-jump of the helicity modulus; for big values $J_{||} \gg A$ the transition should also be of the BKT type.
Then we expect that in the low-temperature (ordered) phase the correlation function $\langle \Delta_r \Delta_{r'} \rangle$ decays algebraically, and in the high-temperature disordered phase it decays exponentially.

\subsubsection{Intermediate $\beta J_{||}$ and the phase diagram}

\label{Intermediate beta Jpar and the phase diagram}

\begin{figure}[tbh]
\centering
  \includegraphics[width=0.8\linewidth]{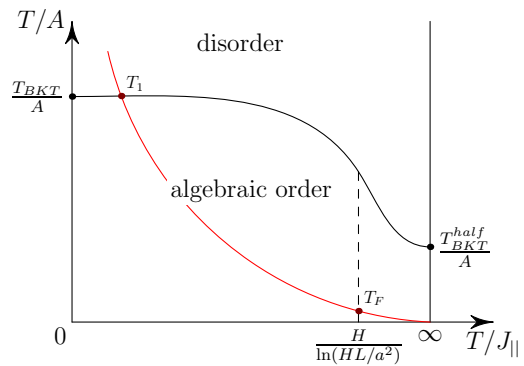}%
\caption{Qualitative phase diagram corresponding to the Hamiltonian (\ref{HamiltonianIsing}) in $T/J_{||}-T/A$ coordinates. The algebraic order area is divided by a crossover (dashed) line into two parts: left, where the large-scale physics is determined by integer vortex-antivortex pairs; and right, where the large-scale physics is determined by half-integer pairs. The red curve shows cooling trajectory, when the number of amplitude solitons is kept fixed.}
\label{Fig3-phase-diag}
\end{figure}

For intermediate $J_{||}$ we expect a crossover 
between the two analyzed limiting cases: strong and weak coupling limits for the Ising subsystem, when integer or half-integer vortices dominate correspondingly. Fig. \ref{Fig3-phase-diag} shows the expected phase diagram of the system.

In the thermodynamic limit, an arbitrarily small $J_{||}$ prevents formation of single half-vortices (Fig. \ref{Fig3-HalfV}).
In the course of heating, the order-disorder transition is of the BKT type with the dissociation of integer vortex-antivortex (VA) pairs. Half-vortex pairs may also be present (Fig. \ref{Fig3-HalfVApair}a), but they are bound by a constant (``confinement'') force, so they always have a finite size and are irrelevant under rescaling.

But if $J_{||}$ is exactly zero, then the confinement force between half-antivortices vanishes. Therefore, they can dissociate, and this indeed takes place at the temperature $T_{BKT}^{half}\approx T_{BKT}/4$.

The crossover between these two regimes ($J_{||} > 0$ and $J_{||} \equiv 0$) occurs at some small $J_{||}$, which is related to the finite size of the system. In order to find this crossover temperature we can recast Kosterlitz-Thouless argument \cite{Kosterlitz:1973}. The free energy of a single half-vortex is
\begin{align}
F = E - TS \simeq J_{||} H - T \ln (H L/a^2)
\label{FreeEnergy}
\end{align}
where $L$ and $H$ are the system dimensions in the chains' and in the transverse directions respectively.

If half-integer VA pairs can escape from the linear confinement at $T<T_{BKT}/4$, then the BKT transition will occur via dissociation of these pairs.
If half-VA pairs escape from the linear confinement at $T_{BKT}/4 < T < T_{BKT}$, then the transition from the algebraically ordered to the disordered state occurs via Ising mechanism (such a possibility was recently explored by Shi et.al.\cite{Shi:2011}).

From (\ref{FreeEnergy}) we find the critical value of $T/J_{||}$, when half-VA pairs become important:
\begin{align}
\frac{T}{J_{||}} = \frac{H}{\ln (H L/a^2)}
\label{TJ}
\end{align}
This crossover temperature (the dashed line in Fig. \ref{Fig3-phase-diag}) goes to infinity as $H\rightarrow\infty$.

\subsection{Pattern formation}

\begin{figure}[tbh]
\centering
\subfloat[]{%
  \includegraphics[width=0.3\linewidth]{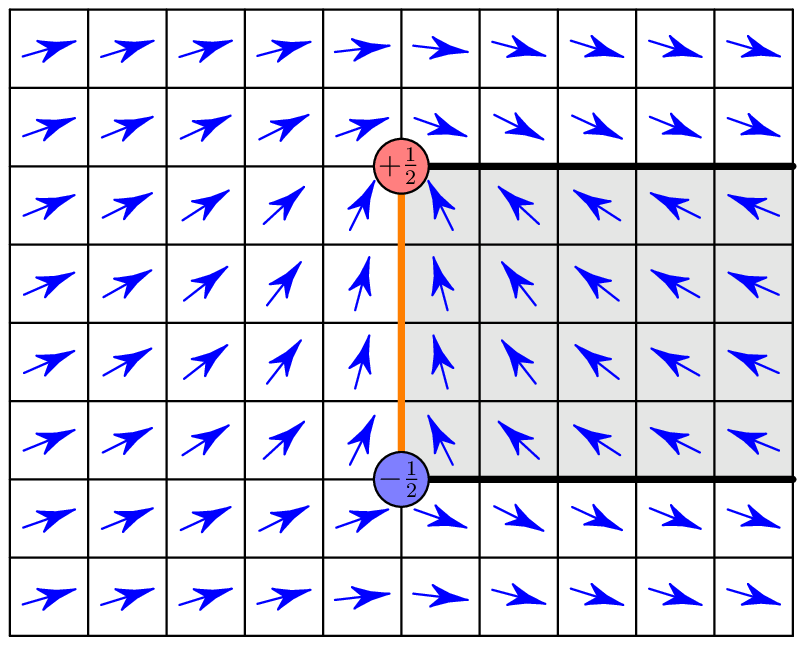}%
}
\quad
\subfloat[]{%
  \includegraphics[width=0.3\linewidth]{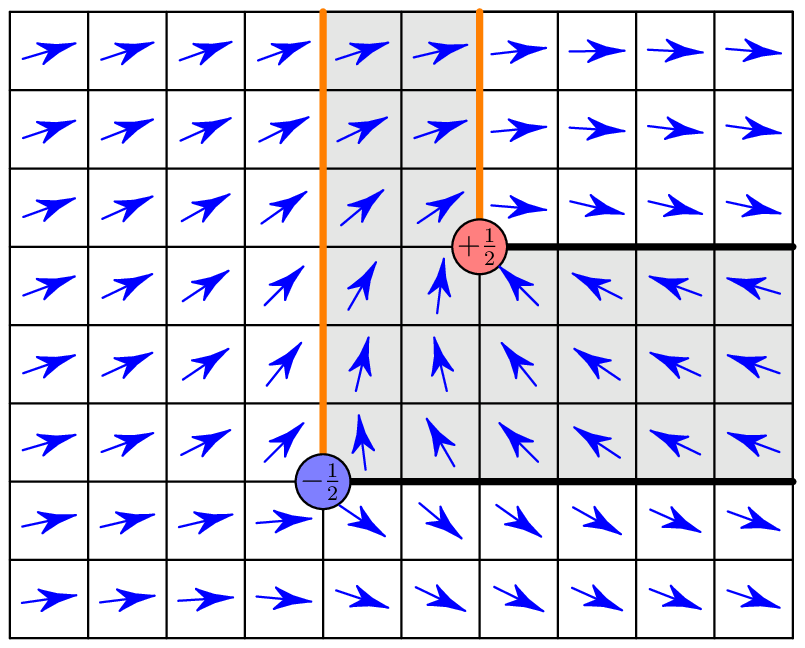}%
}
\quad
\subfloat[]{%
  \includegraphics[width=0.3\linewidth]{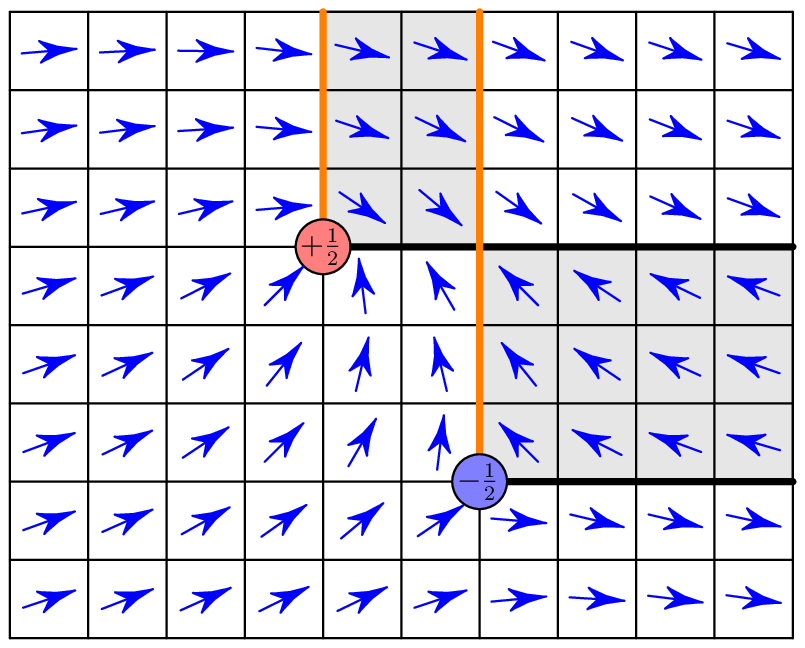}%
}
\caption{Half-integer vortex-antivortex pairs. a) $T>T_2$, the pair structure is dominated by the linear ``confinement'' potential, solitonic walls (orange) are finite. b,c) $T<T_2$, pair structure is dominated by 2D Coulomb log-potential, solitonic walls grow to infinity.}
\label{Fig3-HalfVApair}
\end{figure}

In this subsection we describe the pattern formation in relation to the phase diagram.
Consider the cooling trajectory across the phase diagram when the number of amplitude solitons is fixed (red line in Fig. \ref{Fig3-phase-diag}). In the high-temperature phase, the vortices exist in a plasma state and half-vortices are strongly confined via the linear potential, typically only the smallest half-VA pairs with the transverse distance $1$ are present. At the temperature $T_1$ the system enters the region of algebraic order and the vortices become confined in VA pairs. The earlier free amplitude solitons cannot survive anymore and they start to be dressed by half-VA pairs. Upon further cooling at some crossover temperature $T_2$ (an estimate is given below in (\ref{T2})) the solitons start to aggregate in the transverse direction, thus elongating the half-VA pairs. At some size-dependent temperature $T_F \sim  T_2$, $J_{||}(T_2)$ becomes so small (such that (\ref{TJ}) is fulfilled) that the solitonic walls grow across the whole sample (Fig. \ref{Fig3-HalfVApair}b,c), then the confinement force does not play any further role, half-VA pairs are weakly confined only by 2D Coulomb log-interaction. Since at such low temperatures half-VA pairs are already very rare, then the regular  ``stripe phase'' sets in.

The picture of the two phase transitions resembles the previously studied case of Ising model \cite{Karpov:2016}. For that case, with lowering the temperature, firstly the long-range order (versus quasi-long-range algebraic order described here) is established, and then at the lower temperature $T_2$ the solitonic walls grow across the whole system.

\subsection{Estimates for $T_1, T_2$.}
\label{Subsec_Estimations_for_T1_T2}

In this subsection we find estimates for the transition temperatures $T_1, T_2$.

$T_1$ is the BKT transition temperature, $T_1 \approx T_{BKT} \approx 0.9 A$ [\onlinecite{Shugard:1978}, \onlinecite{Olsson:1995}], and corrections in soliton concentration $\nu$ can be found perturbatively.

$T_2$ can be estimated by extrapolating of $J_{||}(T)$ from high to low temperatures in the approximation of non-interacting solitons (assuming their small concentration: $\nu\ll 1$). Each soliton bears the energy $2 J_{||}$ via Ising subsystem; XY subsystem tries to ``cure'' the defect, then the half-VA pair forms with the minimal transverse size of $1$ site and the energy of the half-vortex cores $2 E_{core} \equiv 2 \tilde{E}_{c} A$ ($\tilde{E}_{c}$ is the dimensionless core energy of a half-vortex). The total energy of such combined defect is $2 J_{||} + 2\tilde{E}_{c} A$. Hence their concentration is
\begin{align}
\nu \approx \exp \left( -\frac{2 J_{||} + 2\tilde{E}_{c} A}{T} \right)
\label{nu}
\end{align}
Inverting this relation for $J_{||}$, we get
\begin{align}
J_{||} (T,\nu) \approx \frac{T}{2} \ln\frac{1}{\nu} - \tilde{E}_{c} A
\end{align}
From the condition $J_{||} (T,\nu) \approx 0$, when the approximation of non-interacting solitons breaks and they start to aggregate in transverse ``rods'', we find an estimation for $T_2$:
\begin{align}
T_2 \approx \frac{2\tilde{E}_{c} A}{\ln (1/\nu)}
\label{T2}
\end{align}

At lower $T \lesssim T_2$ we cannot further neglect the interaction of solitons. Two scenarios may be envisaged then: (i) $T_2$ can be a crossover temperature, below which solitons start to aggregate in transverse lines that are gradually growing as $T\rightarrow 0$; (ii) $T_2$ can be a phase transition temperature, at which infinite lines appear, crossing the whole sample. In the next section we test these two scenarios numerically with results in favor of the scenario (i).

The estimate for $T_2$ can also be done in another way --
using the argument of solitonic transverse ``rods'' \cite{Bohr:1983}. When small transverse aggregates become favorable, they can further stick together, growing into infinite walls. This can be supported by the following argument. Let transverse lines of lengths $l > 2$ became favorable at $T\rightarrow0$. The energy of two such lines is $\sim 2 \ln l$. Merging them into a single line of length $2l$, the energy becomes $\sim \ln(2l) = \ln l + \ln 2$. Therefore for $l>2$ such merging always lowers the energy, which provides the possibility for the rods to grow to infinity.

Here we consider a thermodynamically equilibrium system of such transverse segments of solitonic lines\cite{Bohr:1983} (the solitonic ``rods'').
We shall work with the grand canonical ensemble of solitons with the chemical potential $\mu$ to be determined a posteriori by the fixed average concentration $\nu$. 

Let the amplitude solitons be aggregated in rods of variable lengths $l$. Because of exchange among the rods, their chemical potentials are $\mu(l) = \mu l$. Their energies $E(l)$ can be estimated as:
\begin{align}
E(l) = E_s l + 2 \tilde{E_c} A + \tfrac12 \pi A \ln l.
\end{align}
Here $E_s$ is the core energy of the amplitude soliton; $2 E_c \equiv c A$ is the energy of two half-vortex cores and $\tfrac12 \pi A \ln l$ is the interaction between them.

The distribution of rods over their lengths \cite{Bohr:1983}
\begin{align}
n(l) &= \exp \left( \frac{\mu(l)-E(l)}{T} \right) = \nonumber\\
&=\exp \left( \frac{\mu l - E_s l - 2 \tilde{E}_c A - \tfrac12 \pi A \ln l}{T} \right).
\end{align}
Denote $m = (E_s-\mu)/T > 0$, $\alpha = \pi A/2T$, so that $n(l) = e^{-2 \tilde{E}_c A/T} l^{-\alpha} e^{-ml}$. Then the total concentration of solitons is
\begin{align}
\nu = \sum_{l=1}^{\infty} l \, n(l) &= e^{-2 E_c/T} \sum_{l=1}^{\infty} l^{-\alpha+1} e^{-m l} =\nonumber\\
&=e^{-2 E_c/T} \Li_{\alpha-1} (e^{-m})
\end{align}
where $\Li_s (z) = \sum_{n=1}^{\infty} z^n/n^s$ is the polylogarithm function; its maximum value is reached at $z=1$, when the series is just Riemann zeta function: $\Li_s (1) = \zeta(s)$.

As $T\rightarrow 0$, $e^{-2 \tilde{E}_c A/T} \rightarrow 0$, so to keep $\nu$ fixed, the factor $\Li_{\alpha-1} (e^{-m})$ need to grow to infinity. However for $\alpha>2$ ($T < \pi A /4
\simeq T_{BKT}$) it is bounded from above; therefore the ensemble of finite lines cannot further accommodate all the solitons in the system. This means that infinite (crossing the sample width $H$) domain walls have to grow.

Assuming that the corresponding critical temperature $T_2 \ll T_{BKT} \approx 0.9 A$ (which we then check for self-consistency), so that $\alpha \gg 1$ and $\zeta(\alpha-1) \approx 1$, we get
\begin{align}
\nu \leq e^{-2\tilde{E}_c A/T} \zeta(\alpha-1) \approx e^{-2\tilde{E}_c A/T}
\end{align}
Therefore, the characteristic temperature for transverse walls nucleation is
$T_2 \approx 2\tilde{E}_c A/\ln(1/\nu)$; this estimate is consistent with the one based on the $J_{||}=0$ argument (see (\ref{T2})).
The assumption $T_2 \ll T_{BKT}$ is satisfied when $\nu \ll 1$.

When the temperature decreases towards $T_2$, the solitonic rods begin to elongate, then above some critical length $l^*$ it may become energetically favorable for them to group into pairs and form the bisolitonic rods. In the next section we estimate this characteristic length $l^*$ and discuss how it is affected by the system's anisotropy.

\begin{figure}[tbh]
	\centering
	\subfloat[]{\includegraphics[height=3cm]{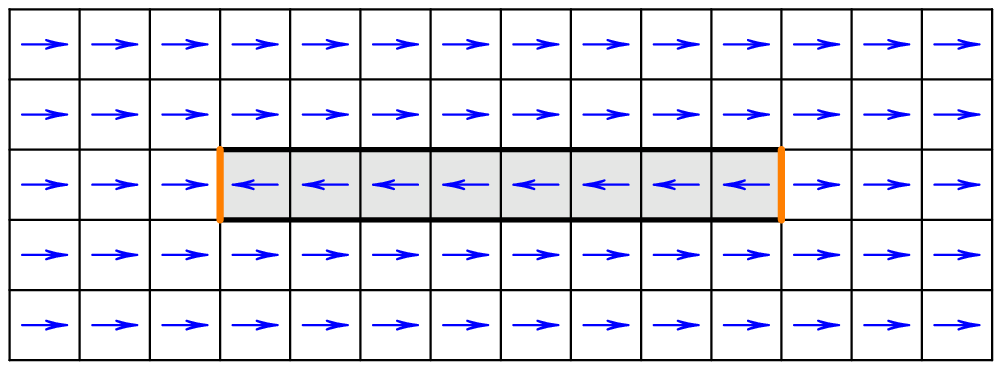}}
	\qquad
	\subfloat[]{\includegraphics[height=3cm]{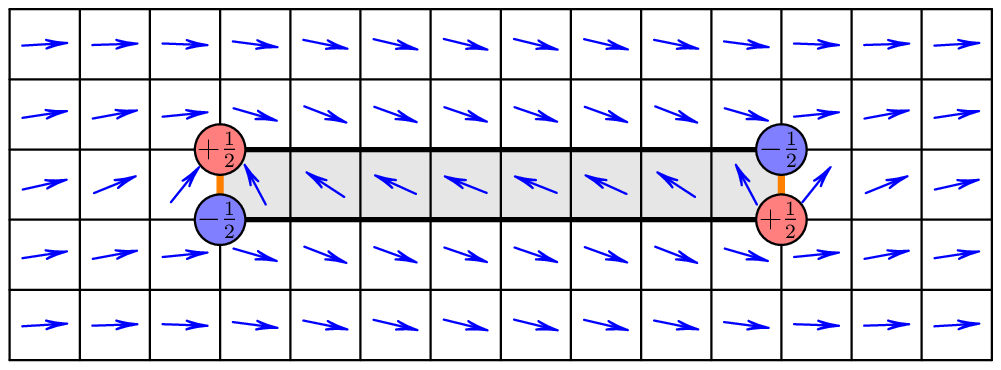}}
	\qquad
	\subfloat[]{\includegraphics[trim={2.85cm 0 3.6cm 0},clip,height=3cm]{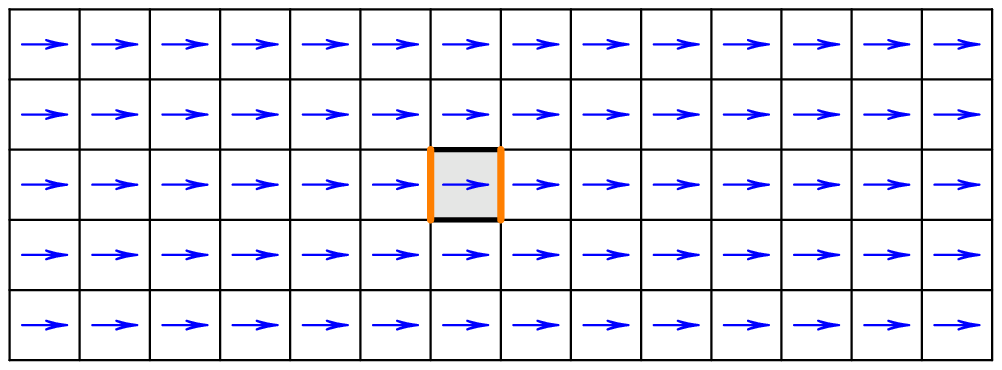}}
	
	\caption{Most energetically favorable configurations for two solitons for cases:  a) $A_{\perp} \gg A_{||}$, b) $A_{\perp} \simeq A_{||}$,  c) $A_{\perp} \ll A_{||}$. In case a) all excess energy is concentrated in vertical orange links, b) the energy excess is smeared around the cores of the half-vortices, c) it is concentrated in black horizontal links.}
	\label{Fig-Anisotropy-solitons}
\end{figure}

\subsection{Role of the anisotropy}

In this subsection we discuss the effect of the anisotropy $A_{\perp}/A_{||} \neq 1$ on the growth of the rods and the $T_2$ transition,  and also find the characteristic length $l^*$, above which solitonic rods combine to bisolitonic ones.

Simple scaling arguments show that BKT-transition temperature $T_1$ depends only on $A=\sqrt{A_{||} A_{\perp}}$ [\onlinecite{Pereira:1995}], so the expression $T_1 \approx 0.9 A$ holds. Likewise, the vortex-antivortex interaction energy also depends only on $A$, therefore the estimate (\ref{T2}) for $T_2$ also holds. However, as we argue below, anisotropy can affect the character of the $T_2$ transition: it can facilitate the growth of either solitonic or bisolitonic rods.

Consider a low-temperature regime of a system with just two solitons (or one bisoliton) confined to a given chain.
Depending on the ratio of $A_{\perp}/A_{||}$ it is energetically favorable for the system to either have separate solitons or a  tight bisoliton (Fig. \ref{Fig-Anisotropy-solitons}). Configuration (a) with two kinks in $\theta$-phase has the energy $4 A_{||} l$, configuration (b) with two half-VA pairs has the energy $(4 \tilde{E}_c + \pi \ln l) \sqrt{A_{||} A_{\perp}}$, and configuration (c) with a tight bisoliton has the energy $4 A_{\perp}$ (for generality we also took into account the transverse length of the rods $l$; in case of Fig.  \ref{Fig-Anisotropy-solitons} we have $l=1$).  Depending on the ratio $A_{\perp}/A_{||}$ one of the three configurations is favored for a given $l$.

Energy of the configuration of Fig. \ref{Fig-Anisotropy-solitons} (a) strongly depends on $l$ and it becomes unstable towards configuration (b) already for rather small values $l \gtrsim \tilde{E}_c \sqrt{A_{\perp}/A_{||}}$. Since the energy of the configuration (b) depends on $l$ only logarithmically, then such type of configurations can be stabilized for a broad range of rods lengths $l<l^* \sim \exp(\frac{4}{\pi} \sqrt{A_{\perp}/A_{||}})$, which exponentially depends on the anisotropy of the system.

This means that for $A_{\perp}/A_{||} \gtrsim \tilde{E}_c^2$, when stable solitonic rods exist, their growth occurs in two stages: first, solitonic rods grow to the characteristic size $l^*$, after which  it becomes favorable to glue two solitonic rods (with the energy $\sim A\ln l$) to one bisolitonic rod with compensating pairs of half-vortices at their terminals; then only the energy  $\sim A_{\perp}$ is left. After this gluing, the system enters the regime of slow further growth of bisolitonic rods, similarly to the previously studied Ising case \cite{Bohr:1983,Teber:2001,Karpov:2016}. This means that here also in the 2D case $T_2$ is only a crossover and not a phase transition temperature.

We note, however, that for non-equilibrium processes, like supercooling the system with a sudden quench (as it is implemented in our Monte Carlo modeling, see Sec. \ref{Subsec Quench}), this first stage can be drastically promoted, since a growing solitonic wall creates uncompensated half-vortices at the ends which attract mobile smaller rods, while the slow process of gluing of low-mobile big rods can be kinetically suppressed.

The above analysis does not take into account the screening effects from multiple interactions of vortices \cite{Kosterlitz:1973}, which limits the literal  applicability to the case of distant and non-interacting solitonic or bisolitonic rods.
That leads to a restriction  upon their characteristic length
$ l^* \ll 1/\nu $ ($l=1/\nu$ is such a length when every square $l\times l$ contains one rod of length $l$, which means that $\nu=l/l^2=1/l$). When the rods become too long i.e. for $l^* \gtrsim 1/\nu$, the interaction among them cannot be further neglected and the system enters the  strongly interacting regime of plasma of half-vortices and antivortices.

We conclude that in 2D the walls' formation temperature $T_2$  is a crossover, becoming for finite samples effectively a phase transition with the transverse size dependent temperature $T_F(H)$. Depending on the ratio of the interchain couplings $A_{\perp}/A_{||}$, either solitonic or bisolitonic walls grow across the sample at $T<T_F(H)$. Contrarily, only topologically trivial bisolitonic walls could be formed for the case of the real order parameter described by the Ising model \cite{Bohr:1983,Teber:2001,Karpov:2016}.

\subsection{Extension to a 3D system}
\label{Subsec Extension to a 3D system}

For a 3D system we expect the phase diagram similar to 2D one (Fig. \ref{Fig3-phase-diag}) except for a character of phase transitions. In the low-temperature ordered phase, there is still no Ising order. This can be seen using the same argument as in Sec. \ref{subsec_strong_coupling}): the combined XY-Ising defect has a finite energy, so the entropy gain in the thermodynamic limit will always favor such defects at all $T>0$, thus breaking the Ising order. This leads to only a ``nematic'' $2\theta$-order in XY subsystem (i.e. nonzero expectation value $\langle \cos(2(\theta_r-\theta_{r'})) \rangle$ opposed to a pure 3D XY model that possesses true long-range $\theta$-order $\langle\cos(\theta_r-\theta_{r'})\rangle\neq 0$ for $|\mathbf{r}-\mathbf{r'}|\rightarrow\infty$).
In terms of the original picture of solitons this means that an isolated soliton with a finite energy is allowed to exist, since the  interchain order which it breaks can be cured by the adjusted phase degree of freedom; such isolated solitons break the Ising order at any non-zero temperature.
This ``nematic'' order can be also described by the combined XY-Ising variable $\Delta_r=  s_r  e^{i\theta_r}$, which acquires a non-zero expectation value below the ordering-transition temperature.

Repeating the arguments of the Sec. \ref{Subsec_Estimations_for_T1_T2}, we obtain the estimate (\ref{T2}) for the walls' formation temperature $T_2$, with a difference that now it is the true transition rather than a crossover.

Instead of the vortex pairs, the lowest energy topological excitations in 3D XY model are the closed vortex loops.
Proliferation of finite vortex loops into infinite ones triggers the order-disorder phase transition \cite{Kleinert:book-multivalued}.

The energy of the loop of radius $R$ is:
\begin{align}
E(R) = \frac{1}{2} \pi^2 A R \ln R + 2\pi R \tilde{E}_c A + \pi R^2 E_s
\label{eq:E_disk}
\end{align}
where the first term describes the loop vorticity energy \cite{Williams:1987}, the second term is the core energy of the vortex loop (distributed along the length of the loop; $\tilde{E}_c$ is a numerical constant, which is generally different from the analogous one in 2D), and the third term is the energy of the enclosed solitonic disk.
The chemical potential of the disk is $\mu(R) = \pi R^2 \mu$,
therefore the concentration of disks with a radius $R$ is:
\begin{align}
n(R) &= \exp \left( \frac{\mu(R)-E(R)}{T} \right)  = \nonumber\\
&= \exp \left( \frac{ (\mu-E_s) \pi R^2 - 2\pi E_c R -  \frac{1}{2} \pi^2 A R\ln R}{T} \right)
\end{align}
Hence the total concentration of solitons is:
\begin{align}
\nu &= \sum_{R=1}^{\infty} \pi R^2 \exp \left( \frac{\mu(R)-E(R)}{T} \right) =\nonumber\\
&= \sum_{R=1}^{\infty} \pi R^2  \exp \left( \frac{ (\mu-E_s) \pi R^2 - 2\pi E_c R -  a A R\ln R}{T} \right)
\label{eq:3D_accumulation_of_solitons}
\end{align}
Clearly, even if $\mu-E_s = 0$, as $T \rightarrow 0$ the sum in the RHS goes to $0$. This means that the disk-phase cannot accommodate all the solitons, and at some $T = T_2$ domain walls need to be produced via a phase transition.

Expressions (\ref{eq:E_disk})-(\ref{eq:3D_accumulation_of_solitons}) hold also for the anisotropic case with $A=A_{\perp}^{2/3} A_{||}^{1/3}$ [\onlinecite{Pereira:1995}].
Analogously to the 2D case the effect of anisotropy can play an important role for the mechanism of growth of the solitonic disks across the sample. If $A_{\perp}/A_{||} \gtrsim \tilde{E}_c^3$, then solitonic disks are stable and can grow up to the characteristic radius $R^*\sim \exp(\frac{4}{\pi^2} \sqrt[3]{A_{\perp}/A_{||}})$, after which they glue to bisolitonic disks.


\section{Numerical simulations}
\label{Sec Numerical simulations}
	
In this section we study numerically the phase transitions, described in the previous section qualitatively.
We model the cooling behavior of a system described by the Hamiltonian (\ref{HamiltonianIsing}) using Metropolis Monte Carlo (MC) method.
The numerical algorithm preserves the number of amplitude solitons (so the term $J_{||} \sum_r s_r s_{r+\hat{x}} = const$, where $r$ runs through all sites of the lattice)  which substantially simplifies the analysis. There is no restrictions upon the phase degree of freedom $\theta$.
An ambiguity of the visual representation related to the gauge transformation (\ref{eq:gauge_transformation}) is resolved by maximizing the amount of $s=+1$ Ising spins when drawing the figures.

\subsection{Observed topological defects}

\begin{figure}[tbh]
\includegraphics[height=3cm]{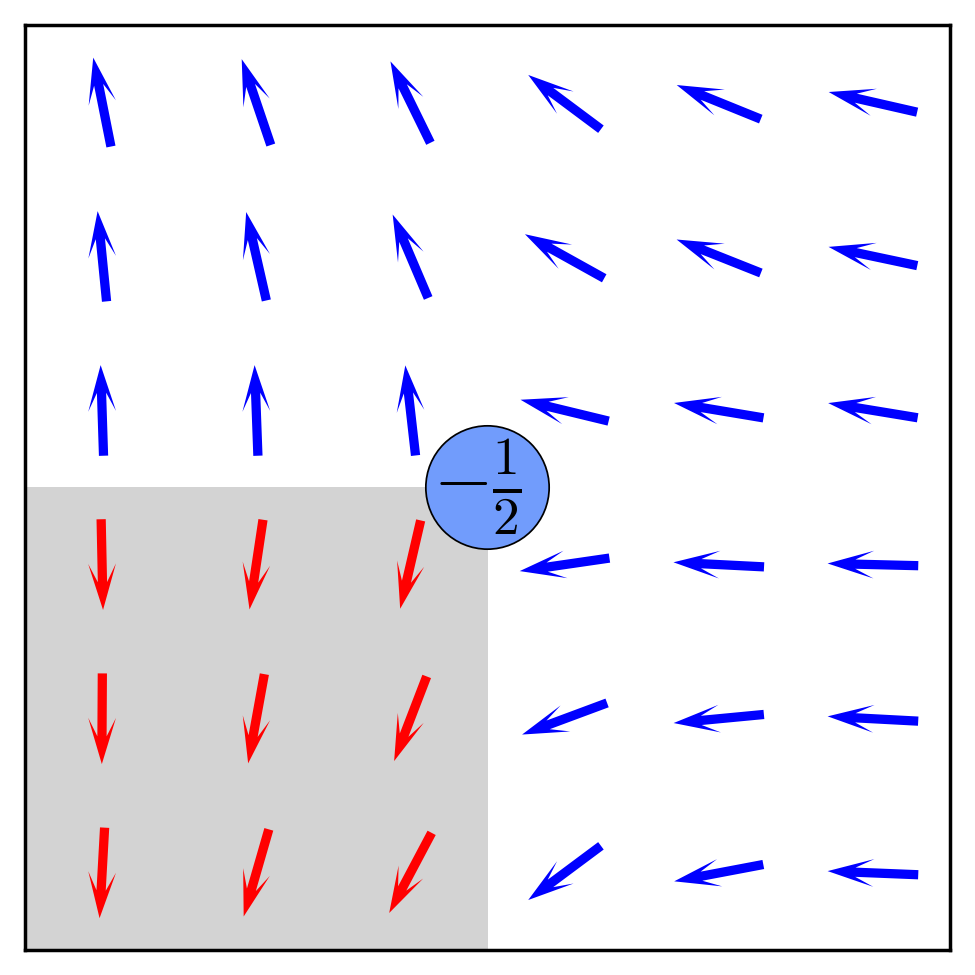}
\caption{A single half-integer antivortex observed in simulations.
(the full image is shown in Fig. \ref{fig_A5}c).
Arrows  show the XY degree of freedom, the gray/white shading shows the Ising degree of freedom. The string of solitons is the boundary separating the differently shaded areas in the chains' ($x$) direction. The blue disk denotes the core of the half-integer antivortex.}
\label{fig_top_defects_antivortex}
\end{figure}

We start with the description of topological defects observed in the modeling.
Figure \ref{fig_top_defects_antivortex} shows the half-vortex enforced by the termination of the line of amplitude solitons (the unabridged version of the  image is shown in Fig. \ref{fig_A5}d). Traveling counterclockwise around the core over the complete $2\pi$-revolution, the arrow (corresponding to the XY degree of freedom $\theta$) rotates by $\pi$ in the clockwise direction -- in some similarity to the field of directors in nematic liquid crystals.

\begin{figure}[tbh]
\includegraphics[width=0.9\linewidth]{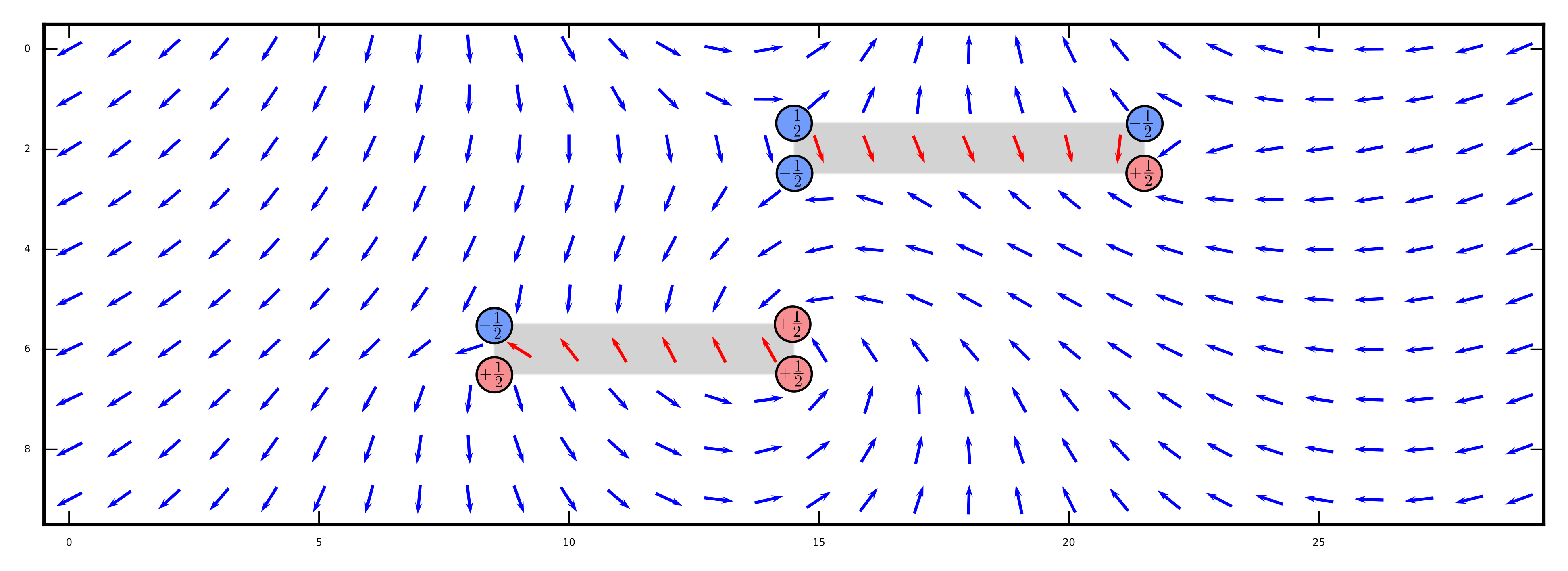}
\caption{Configurations with 4 solitons for a system $10\times30$ with $A_{\perp}=0.25A_{||}$, which was rapidly quenched down to $T \ll A$.
Arrows  show the XY degree of freedom, the shading shows the Ising degree of freedom. The soliton is the boundary separating the differently shaded areas in the chains' ($x$) direction. Red and blue disks denote half-integer vortices and antivortices.}
\label{fig_top_defects_long}
\end{figure}

Figure \ref{fig_top_defects_long} shows a configuration with 4 solitons (i.e. $s\rightarrow-s$ kinks breaking the interchain order) which are ``dressed'' by the phase adjustments $\theta\rightarrow\theta+\pi$, restoring the interchain order. Here we observe two neutral half-VA pairs (outer ones) and also two charged half vortex-vortex and antivortex-antivortex pairs (inner ones). The presence of such configurations with combined phase-amplitude topological defects grows with the increase of $A_{\perp}$. The sharper versions of such combined defects can be also found in a great number in Fig. \ref{fig_A5}.

\begin{figure}[tbh]
\subfloat[]{\includegraphics[width=.49\linewidth]{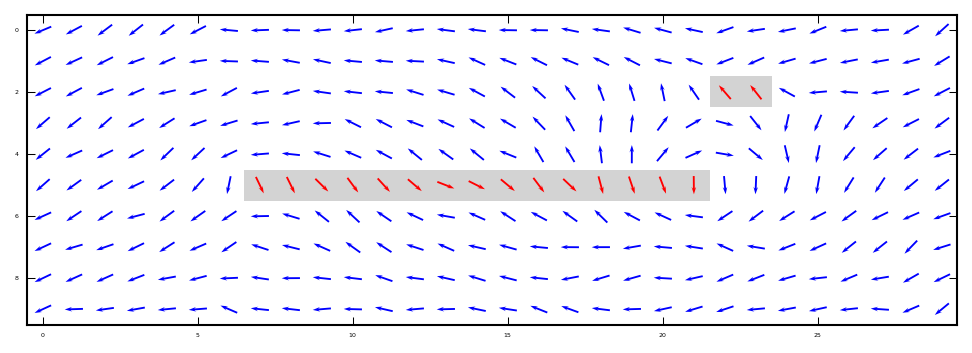}}
\hfill
\subfloat[]{\includegraphics[width=.49\linewidth]{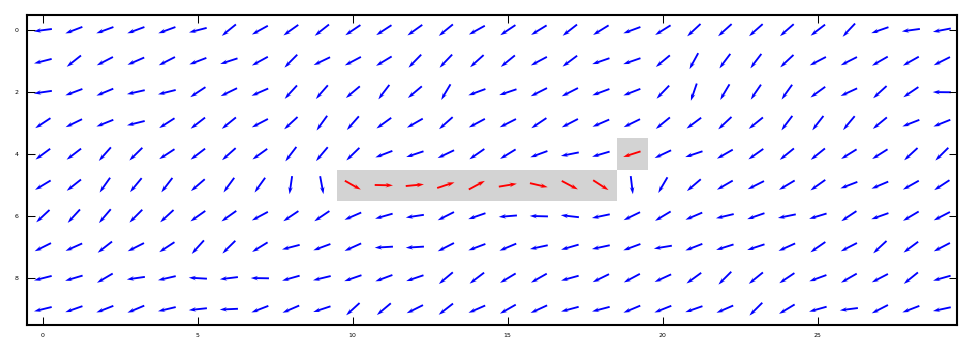}}
\hfill
\subfloat[]{\includegraphics[width=.49\linewidth]{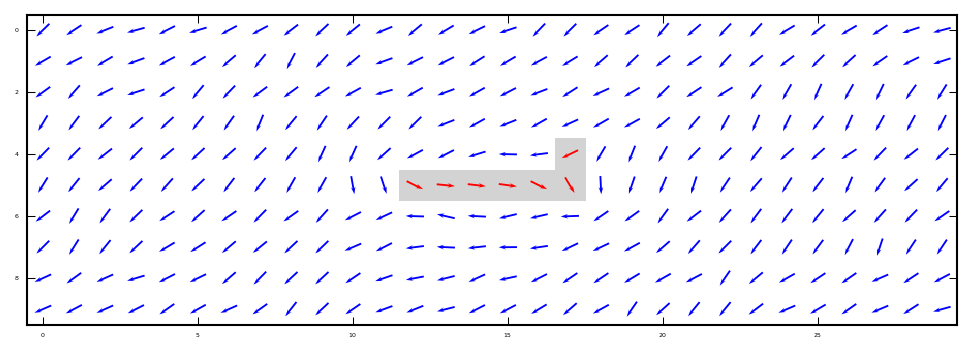}}
\hfill
\subfloat[]{\includegraphics[width=.49\linewidth]{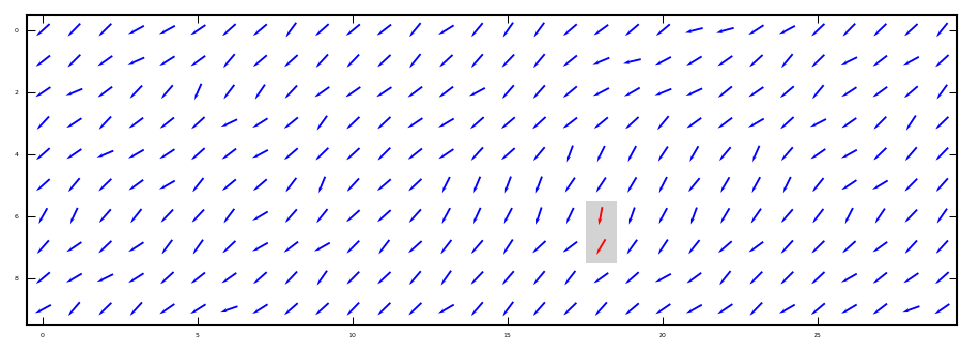}}
\caption{Four snapshots of the MC evolution of the system under the slower cooling. The system with 4 solitons, size $10\times30$,  and $A_{\perp}=0.25A_{||}$.}
\label{fig_top_defects_short}
\end{figure}

Figure \ref{fig_top_defects_short} shows several snapshots of the MC evolution of the system. From (a) to (b) we see the attraction of vortex-vortex to antivortex-antivortex pair and their annihilation with the formation of a segment of solitonic wall. From (b) to (c,d) we observe how the long bisolitonic pair shrinks, presumably because of the weak dipole-dipole interaction of the half-VA pairs at its ends. The final configuration (d) constitutes a segment of a bisolitonic wall, which
no longer possesses a topological defect in $\theta$, with all the energy loss $\sim A_{\perp}$ being concentrated in two transverse $s$-misfits (coming from the term $-A_{\perp} s_r s_{r+\hat{y}}\cos(\theta_r-\theta_{r+\hat{y}})$ in (\ref{HamiltonianIsing})).
The tendency to formation of such bisoliton walls becomes more pronounced with decreasing of $A_{\perp}$ (see also Fig. \ref{fig_A01}).

\subsection{Crossover and finite-size phase transitions in 2D}

Here we study in detail the limiting cases of strong ($A_{\perp}=4 A_{||}$) and weak  ($A_{\perp}= A_{||}/4$) interchain couplings, and also the intermediate symmetric case  ($A_{\perp}=A_{||}$).
In the thermodynamic limit, the properties of these three cases qualitatively match each other, but for finite systems there is a difference among them as described below. For all three cases we take the same value of $A=\sqrt{A_{\perp} A_{||}}$. Concentration of solitons is taken to be $\nu=0.05$ (per site) throughout this subsection.
The size $L\times H$ of the studied system is $25\times100$; periodic boundary conditions are imposed.
The system is cooled down from $T=1.5 A > T_{BKT}$ to $T=0.01A$.

\subsubsection{Strong interchain coupling: $A_{\perp}=2A$, $A_{||}=A/2$.}

\begin{figure}[tbh]
\subfloat[$T=0.31 A$]{\includegraphics[width=1.0\linewidth]{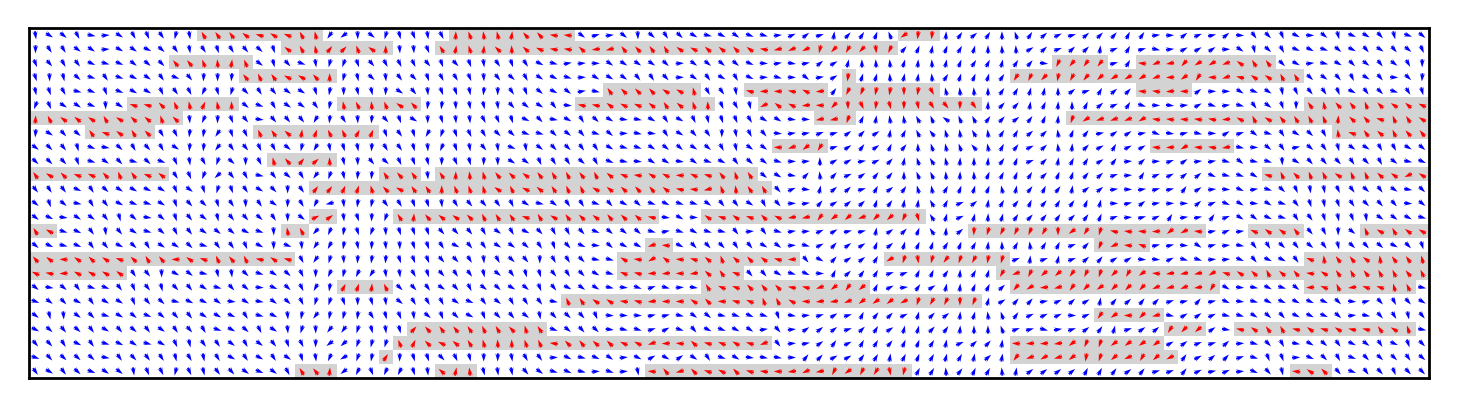}}
\hfill
\subfloat[$T=0.30 A$]{\includegraphics[width=1.0\linewidth]{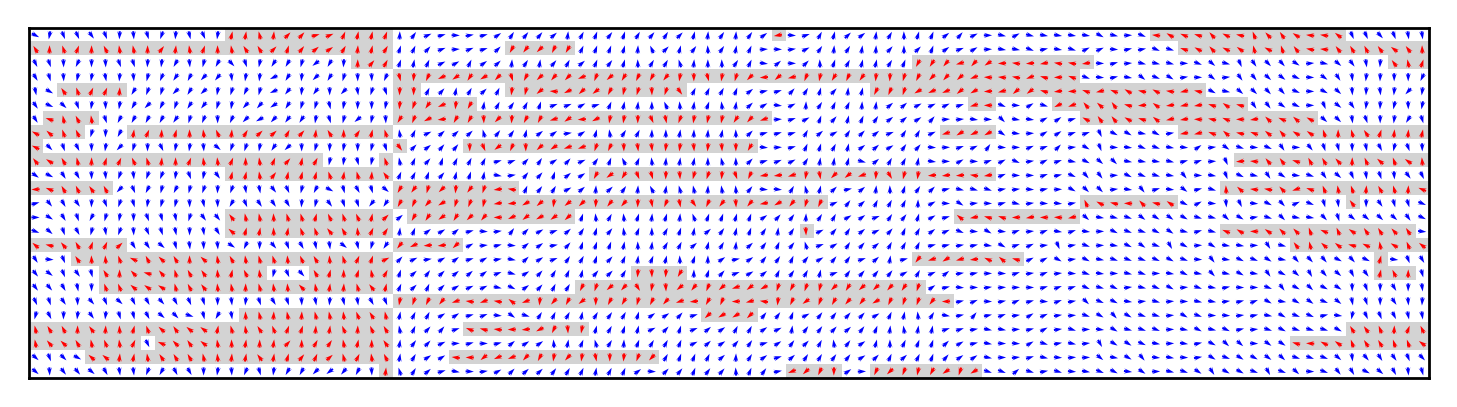}}
\hfill
\subfloat[$T=0.01 A$]{\includegraphics[width=1.0\linewidth]{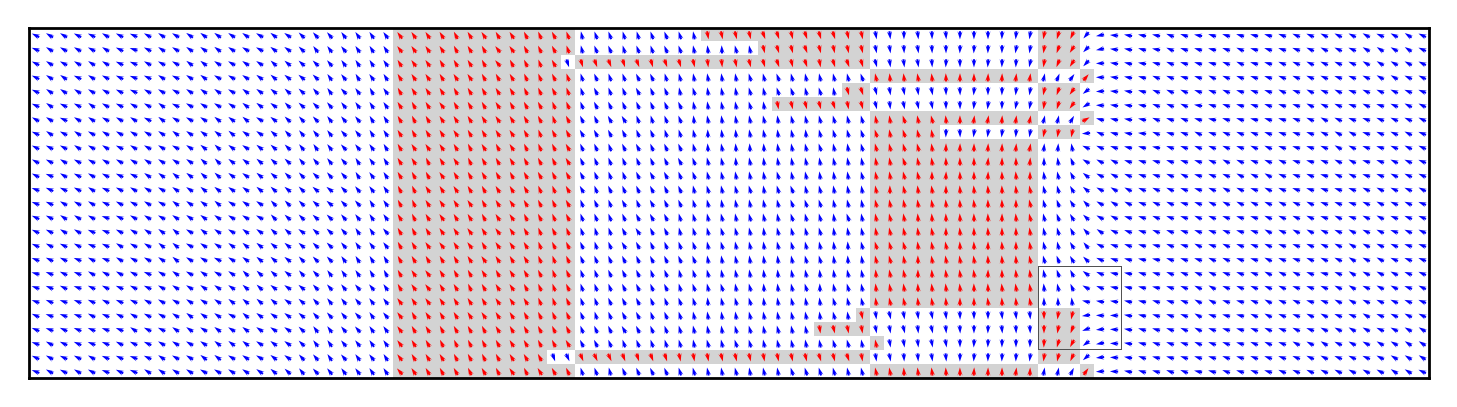}}
\caption{Configurations upon cooling for a system $25\times 100$ with a strong interchain coupling $A_{\perp}=4 A_{||}$. From (a) to (b) we see that the transverse rod of solitons grows across the whole sample.
In (c) the box shows the half-antivortex, presented in Fig. \ref{fig_top_defects_antivortex}.
Arrows' directions show the XY degree of freedom, the shading shows the Ising degree of freedom. The soliton is the boundary separating the differently shaded areas in the chains' $(x)$ direction.}
\label{fig_A5}
\end{figure}
\begin{figure}[tbh]
\centering
  \includegraphics[width=1.0\linewidth]{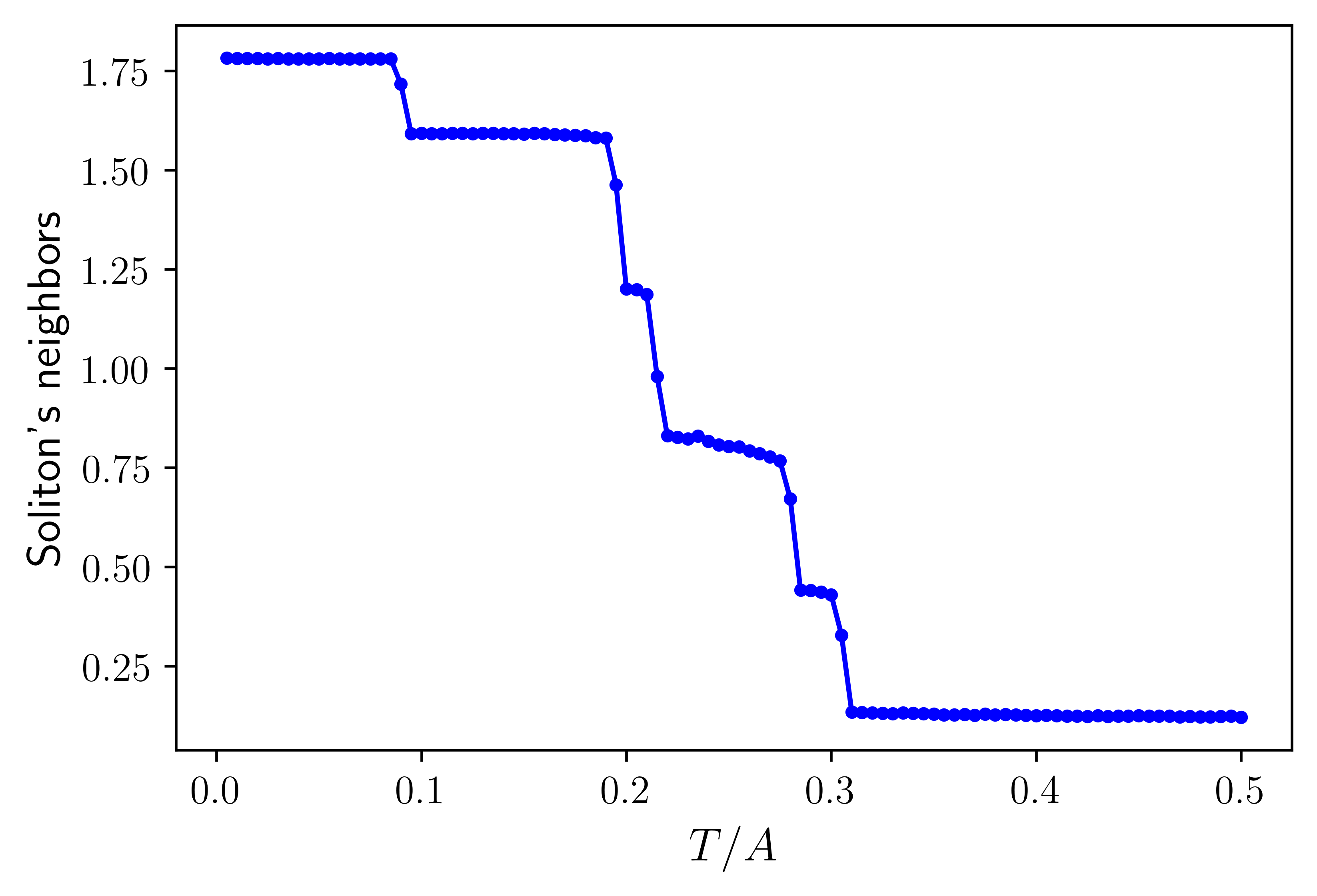}%
\caption{The average number of transverse neighbors of solitons vs temperature (for a system $25\times100$ with a strong interchain coupling $A_{\perp}=4 A_{||}$). Step-like behaviour indicates the sequence of walls installed across the sample width.}
\label{fig_plot_A5}
\end{figure}

We start from the strong interchain coupling: $A_{\perp} = 4 A_{||}$. Cooling the system down starting from the high temperatures, first we observe only short transverse solitonic rods (Fig. \ref{fig_A5}a).
At temperature $T_2\approx 0.3A$ we observe a sudden change: one of the solitonic rods starts to grow until it reaches the borders of the sample (Fig. \ref{fig_A5}b).
With further cooling more solitonic walls appear (Fig. \ref{fig_A5}c) until the material for their building -- free solitons -- is exhausted. All the walls are installed through the grows of solitonic rods.

The transition can also be detected using integrated characteristics. Fig. \ref{fig_plot_A5} shows the dependence of the average number of solitons' transverse neighbors versus temperature. With lowering the temperature, the condensation of the first wall occurs at $T\simeq 0.3 A$. With further cooling, more walls condense sequentially giving rise to the step-like features on the plot.

\subsubsection{Weak interchain coupling: $A_{\perp}=A/2$, $A_{||}=2A$.}

\begin{figure}[tbh]
\subfloat[$T=0.33 A$]{\includegraphics[width=1.0\linewidth]{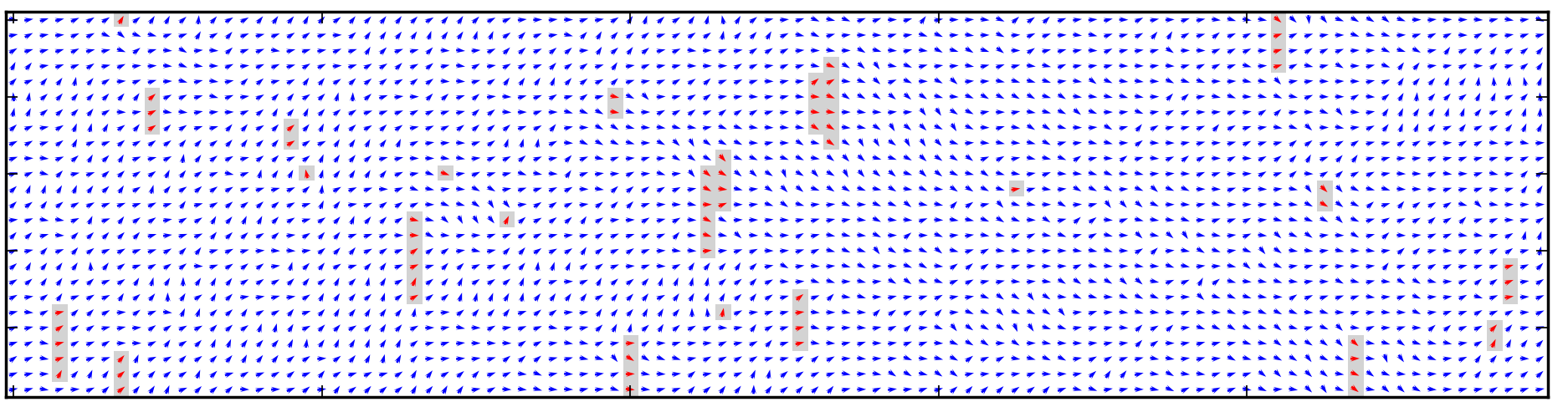}}
\hfill
\subfloat[$T=0.32 A$]{\includegraphics[width=1.0\linewidth]{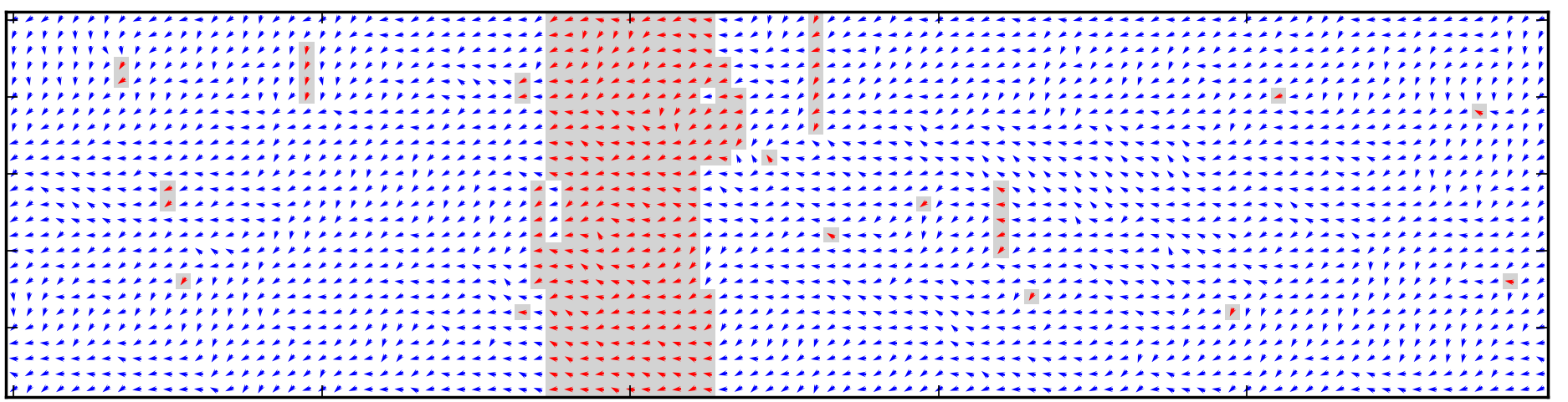}}
\hfill
\subfloat[$T=0.01 A$]{\includegraphics[width=1.0\linewidth]{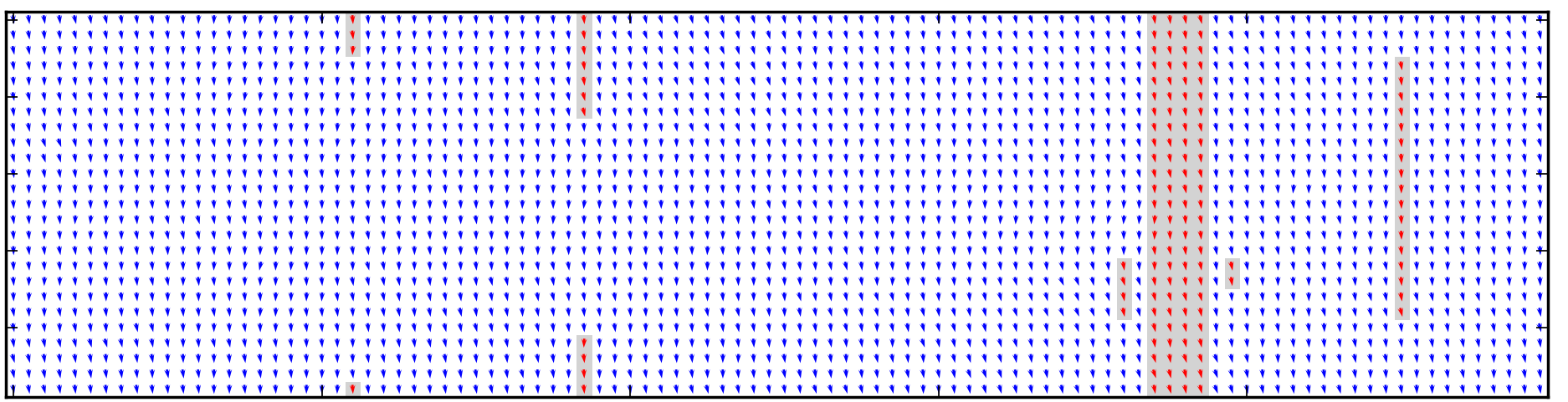}}
\caption{Configurations upon cooling for weak interchain coupling $A_{\perp}=0.25 A_{||}$.}
\label{fig_A01}
\end{figure}
\begin{figure}[tbh]
\centering
  \includegraphics[width=1.0\linewidth]{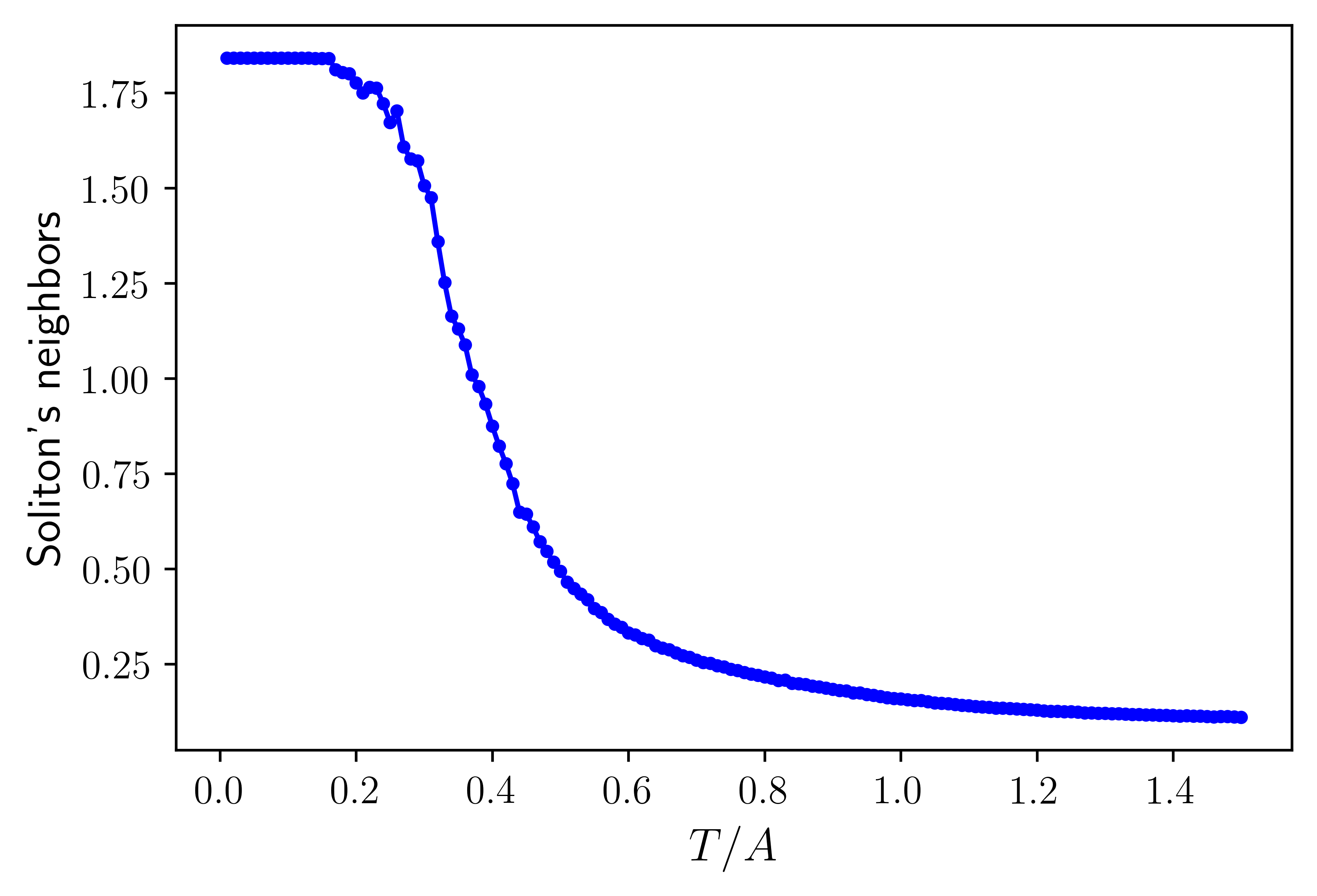}%
\caption{The average number of transverse neighbors of solitons vs temperature (for a system $25\times100$ with weak  interchain coupling $A_{\perp}=0.25 A_{||}$). The finite-size wall-formation crossover happens at $T\sim0.4 A$.}
\label{fig_plotA01}
\end{figure}

Here we consider the case of weak interchain coupling $A_{\perp} = 0.25 A_{||}$. The walls' formation here is similar to the previously studied case of the pure Ising model \cite{Bohr:1983,Teber:2001,Karpov:2016}. At $T\gg A_{\perp}$ different chains are only weakly correlated with each other. 
For $T \sim A_{\perp} \ll A_{||}$ the in-chain variation of the phase is strongly suppressed, so the phase degree of freedom cannot any more ``cure'' the intrachain misfit created by the soliton, thus a (pseudo-) confinement force between solitons is established, which binds them into pairs. 
With further cooling, bisolitons aggregate into growing rods which finally cross the whole sample; after that the paired bisolitonic walls can freely diverge to separated solitonic ones (Fig. \ref{fig_A01}).

Fig. \ref{fig_plotA01} shows the dependence of the average number of solitons' transverse neighbors versus temperature: after the first bisolitonic rod crossing the sample, the fast growth of the number of neighbors with lowering the temperature (starting at $T\sim 0.4A$) finally saturates to almost flat behaviour at $T\approx 0.25 A$, since the system runs out of the new building blocks (free bisolitons) for the further growth of the remaining rods.

\subsubsection{Intermediate interchain coupling: $A_{\perp}=A$, $A_{||}=A$.}

The symmetric case $A_{\perp}=A_{||}$ is somewhat intermediate between the cases of strong and weak interchain couplings and combines some traits of both.

\begin{figure}[tbh]
	\subfloat[$T=0.42 A$]{\includegraphics[width=1.0\linewidth]{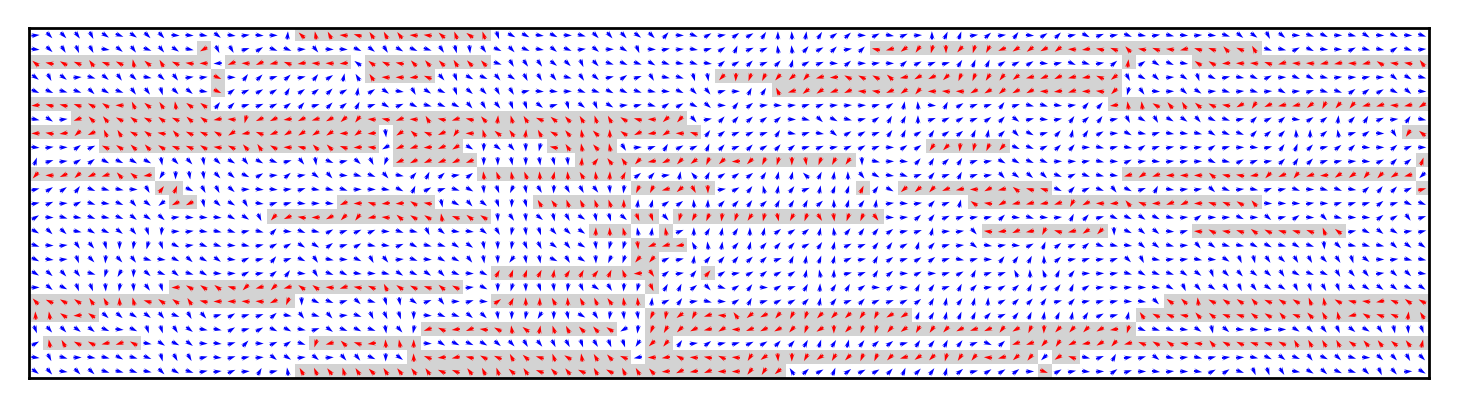}}
	\hfill
	\subfloat[$T=0.40 A$]{\includegraphics[width=1.0\linewidth]{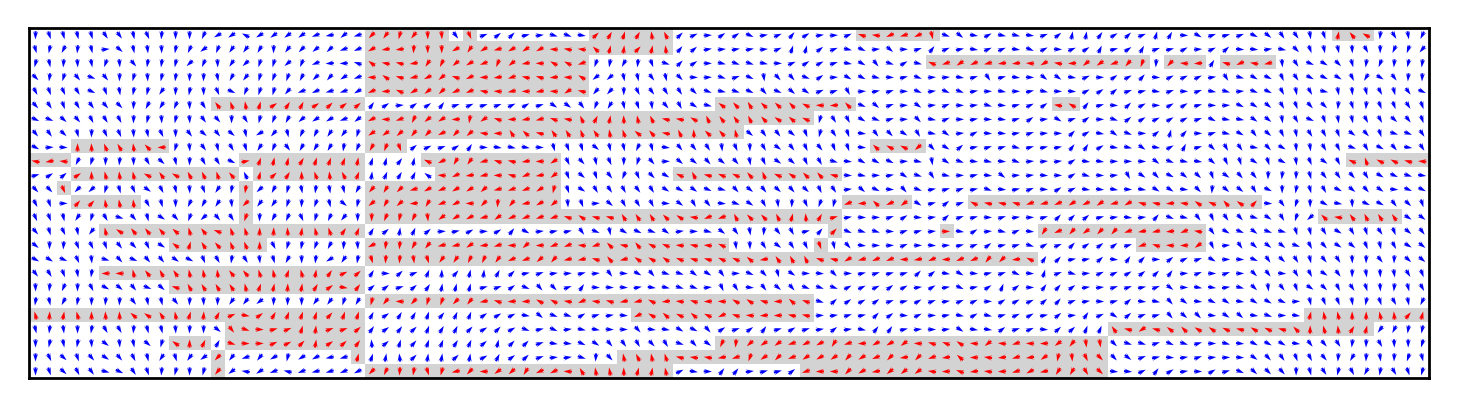}}
	\hfill
	\subfloat[$T=0.01 A$]{\includegraphics[width=1.0\linewidth]{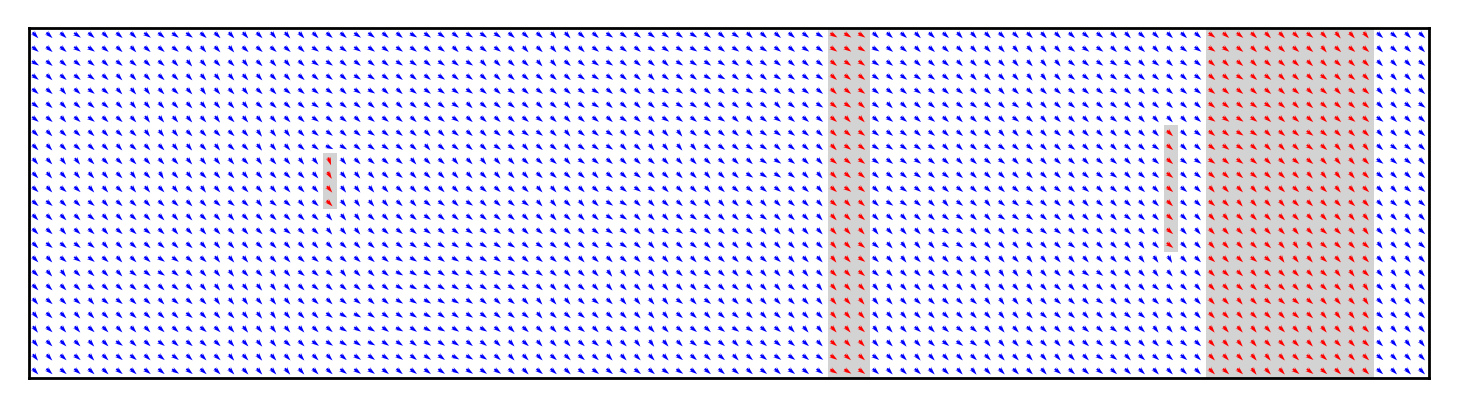}}
	\caption{Configurations upon cooling for the intermediate interchain coupling $A_{\perp}= A_{||}$.}
	\label{fig_A1}
\end{figure}

\begin{figure}[tbh]
	\centering
	\includegraphics[width=1.0\linewidth]{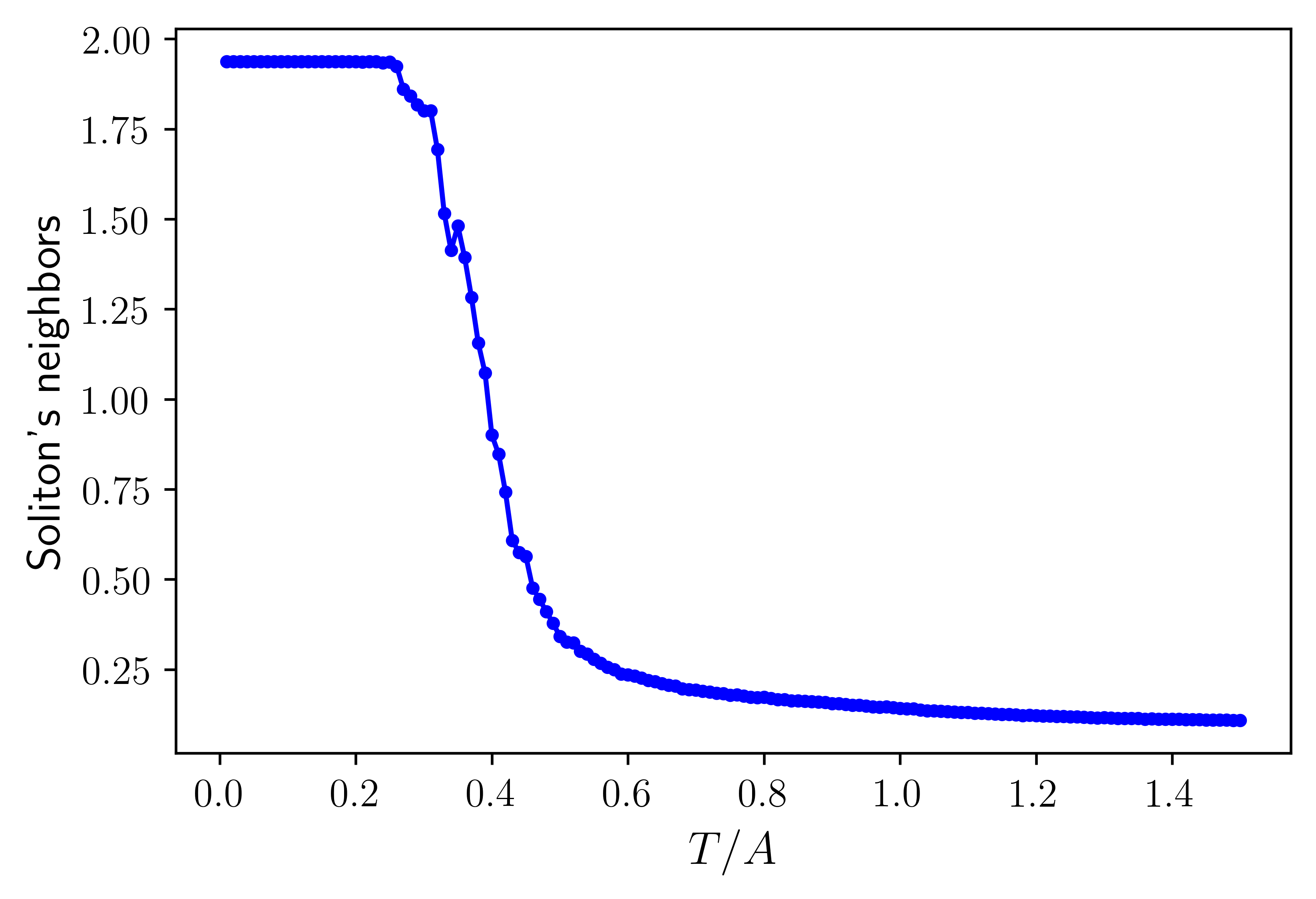}%
	\caption{The average number of transverse neighbors of solitons vs temperature (for a system $25\times100$ with the intermediate interchain coupling $A_{\perp}=A_{||}$). The finite-size wall-formation crossover happens at $T\sim0.4 A$.}
	\label{fig_plotA1}
\end{figure}

Unlike the case of strong interchain coupling, there are stable solitonic rods with transverse lengths down to the smallest size $l=1$ dressed by half-VA pairs, thus solitonic rods elongate gradually starting from the smallest ones. When $T$ approaches $T_2 \approx 0.41A$ from above, there is already a considerable amount of solitonic rods with $l>1$ (Fig. \ref{fig_A1}a).
First walls grow through solitonic rods (Fig. \ref{fig_A1}a), but then the system starts to run out of the solitons, hence the screening of vortices becomes inefficient and solitonic rods recombine to bisolitonic ones, so the rest of the walls grows through the bisolitonic rod mechanism (Fig. \ref{fig_A1}b). The temperature dependence of the number of the transverse neighbors of solitons (Fig. \ref{fig_plotA1}) is qualitatively similar to the case of the weak interchain coupling (Fig. \ref{fig_plot_A5}), which reflects the gradual rods' elongation.


We conclude this subsection by noting that all the cases of weak, intermediate, and strong couplings probably match each other in the thermodynamic limit. Supplementary Figure 2 (Appendix B) shows the results of the simulations for the system $100\times100$ for the case of strong coupling. We observe that solitonic rods can be favorable up to some maximal length, however the domain walls grow through the bisolitonic mechanism. This suggests that even in the case of strong interchain coupling in 2D the walls' growth proceeds through a crossover and not a phase transition.

\subsection{Phase transitions in 3D}

In this subsection we describe results of the numerical modeling for the three-dimensional case. Here we consider the system with the size $10\times10\times100$ with concentration of solitons $\nu=0.06$ and the symmetric coupling $A_{\perp}=A_{||}\equiv A$. System is cooled  from $T=3A$  down to $T=0.02A$.

\begin{figure}[tbh]
	\centering
	\includegraphics[width=1.0\linewidth]{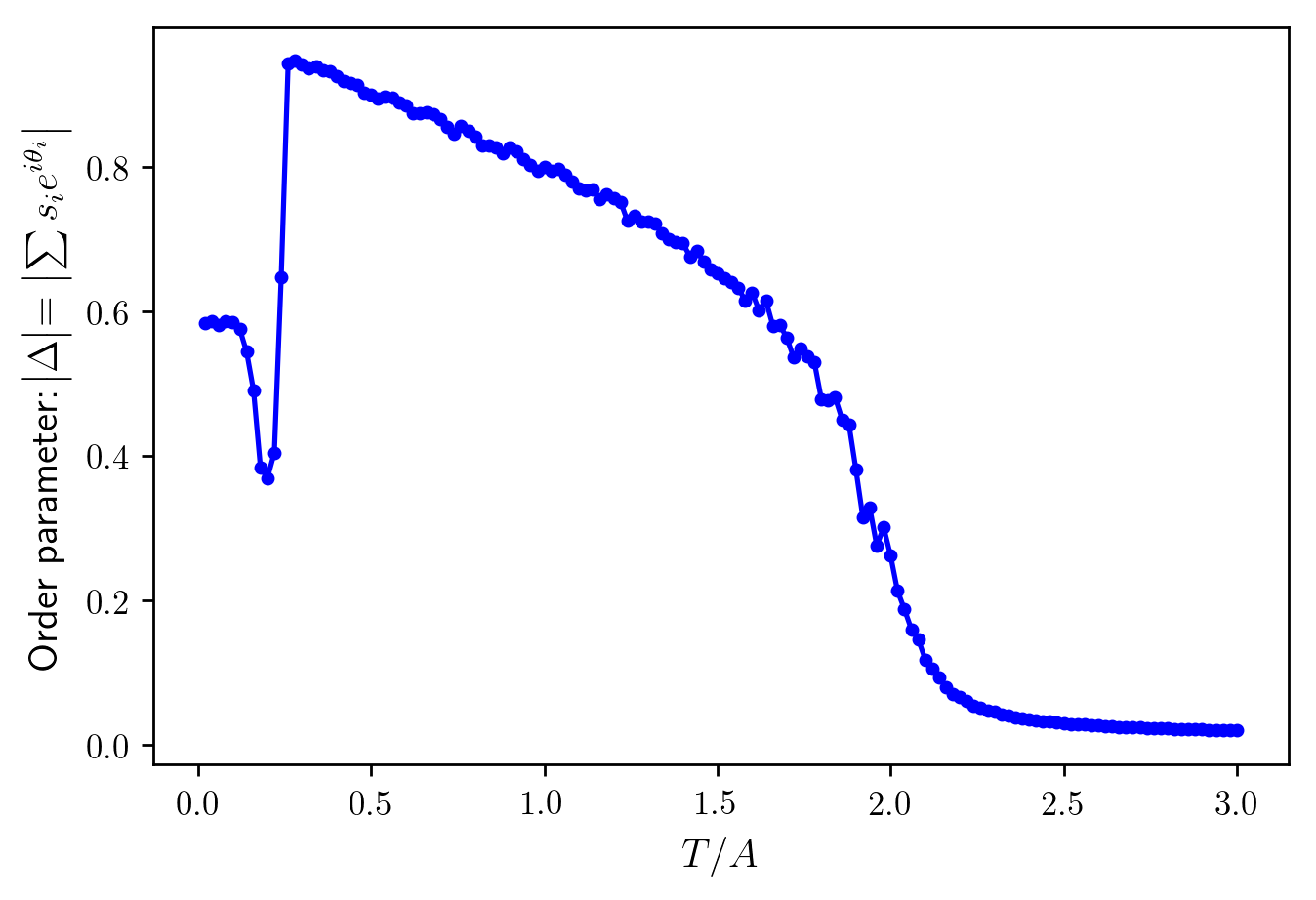}%
	\caption{Temperature dependence of the modulus of 3D complex order parameter $|\Delta|=\left|\sum s_i e^{i\theta_i}\right|$ (for a system $10\times10\times100$ with $A_{\perp}=A_{||}$).}
	\label{fig_plot3D_Order_intermediate}
\end{figure}

\begin{figure}[tbh]
	\subfloat[$T=0.28 A$]{\includegraphics[width=1.0\linewidth]{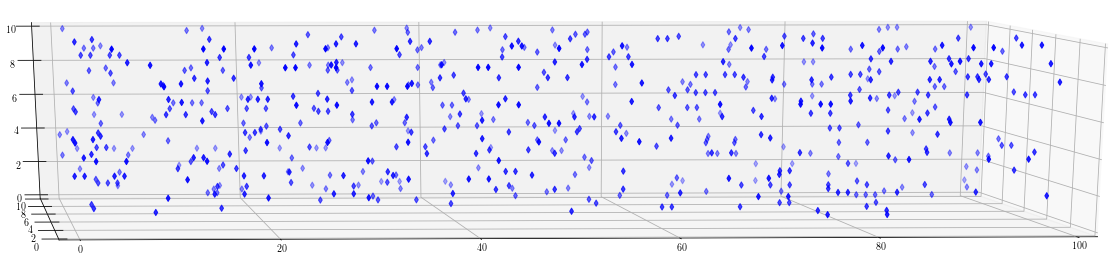}}
	\hfill
	\subfloat[$T=0.24 A$]{\includegraphics[width=1.0\linewidth]{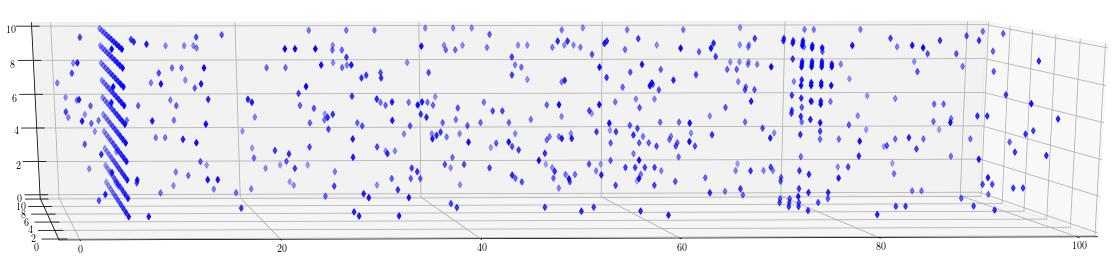}}
	\hfill
	\subfloat[$T=0.02 A$]{\includegraphics[width=1.0\linewidth]{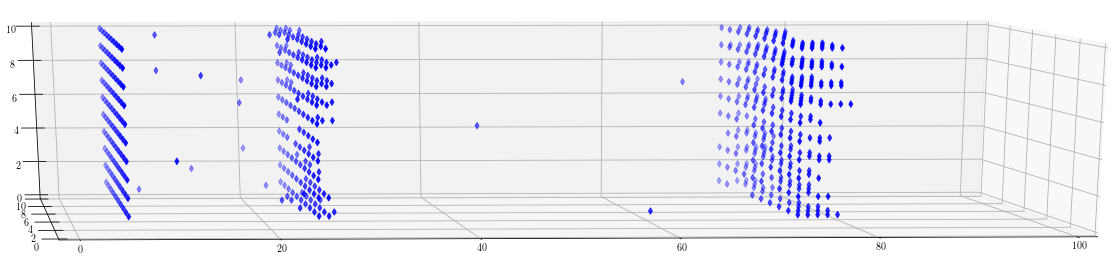}}
	\caption{Configurations upon cooling for a system $10\times 10 \times 100$ with a strong interchain coupling $A_{\perp}=A_{||}$. From (a) to (b) we see that the transverse rod of solitons grows across the whole sample. Blue symbols represent the solitons.}
	\label{fig_3D_intermediate}
\end{figure}

In the three-dimensional case, unlike 2D, there is a true long-range ordering phase transition happening at $T_1 \approx 2A$ (see Fig. \ref{fig_plot3D_Order_intermediate}; the small concentration of solitons only slightly shifts down the 3D XY model transition temperature $T_{XY}^{3D} \approx 2.2A$ [\onlinecite{Pereira:1995}]). With lowering the temperature, the first solitonic wall grows across the sample at $T\approx 0.24 A$. For the periodic boundary conditions used in the simulation, actually, a pair of solitonic walls has to grow -- in Fig.  \ref{fig_3D_intermediate}b we can see the fully grown wall at the left side of the figure and an emerging wall at the right side. The system decomposes to domains with alternating values of the order parameter  $\Delta=\sum s_i e^{i\theta_i}$. In a macroscopic sample we expect an equal amount of $+\Delta$ and $-\Delta$ domains, so the combined XY-Ising order parameter $\Delta$ should vanish; for a finite system it just drops to a smaller value. Still, the system possesses a non-zero value of the nematic order parameter $\langle \cos(2\theta) \rangle$. With lowering the temperature more walls are installed through the bisolitonic mechanism -- Fig.  \ref{fig_3D_intermediate}c.

\begin{figure}[tbh]
	\centering
	\includegraphics[width=1.0\linewidth]{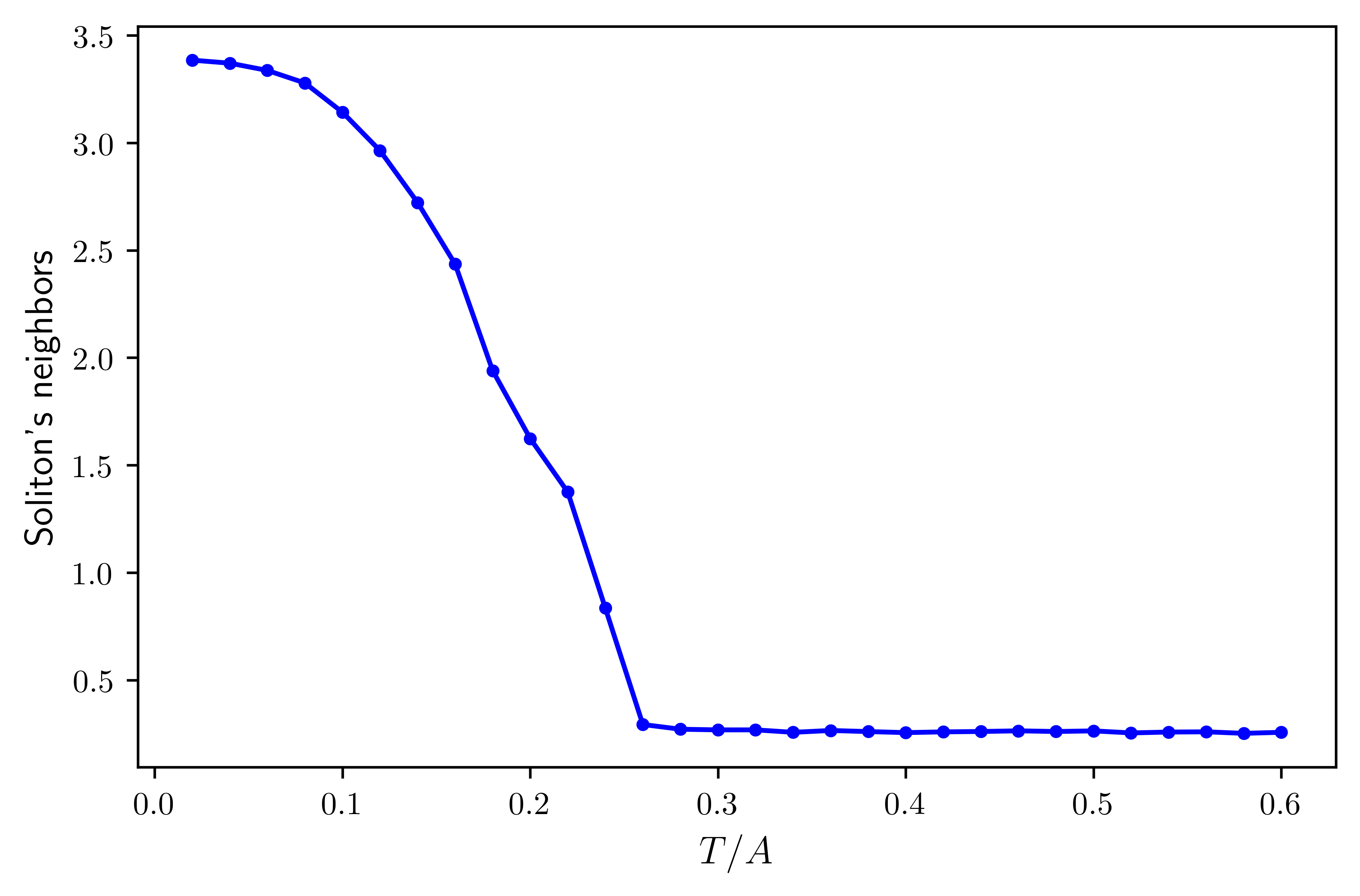}%
	\caption{The average number of transverse neighbors of solitons vs temperature (for a system $10\times10\times100$ with  $A_{\perp}=A_{||}$).}
	\label{fig_plot3D_Neighb_intermediate}
\end{figure}

Temperature dependence of the average number of transverse solitons' neighbors (Fig. \ref{fig_plot3D_Neighb_intermediate}) indicates that there is a pronounced start of the transverse aggregation of solitons at $T\approx 0.26A$, which eventually leads to solitonic walls' formation.

\subsection{Quench}
\label{Subsec Quench}

\begin{figure}[tbh]
	\centering
	\includegraphics[width=1.0\linewidth]{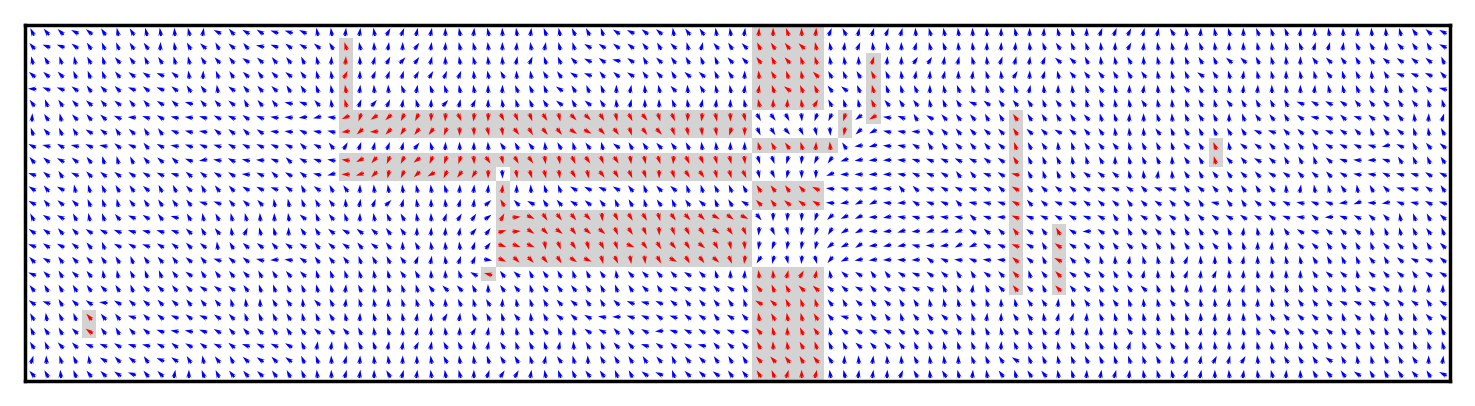}%
	\caption{Configuration of the system after sudden quench from $T>T_2$ to $T=0.25 A < T_2$ (for a system $25\times100$ with  $A_{\perp} = A_{||}/4$).}
	\label{fig_plot2D_supercooling}
\end{figure}

In this subsection we analyze the scenario of a sudden quench from high to low temperatures $T<T_2$.
Such a scenario is relevant to photo- and electrical pulse induced phase transitions \cite{PIPT-book:2004,impact:12}
and in particular to formation of domain walls in CDW systems \cite{Stojchevska:2014,Yoshida:2014,Vaskivskyi:2016,Ma:2016,Cho:2016,Svetin:2016,Gerasimenko:2017,Karpov:2018}.
We show that even for the case of weak interchain coupling the solitonic mechanism of domain walls growth can be important.

Figure \ref{fig_plot2D_supercooling} shows configuration of the system after a sudden quench from a high temperature $T>T_2$ to low temperature $T=0.25 A<T_2$ for a system with weak interchain coupling $A_{\perp} = A_{||}/4$. We observe that the domain wall was created by the solitonic mechanism.
When the solitonic rod initially starts to grow, half-vortices with opposite charges (vorticities) are created at its ends. These charges attract smaller solitonic rods via logarithmic interaction mediated by the XY subsystem,
promoting the further growth of the solitonic rods.
On the contrary, the growth of bisolitonic rods goes via attaching of new bisolitons at the ends, which is essentially a much slower random walk process.
At the same time, longer solitonic rods become low-mobile, therefore their recombination to bisolitonic rods is kinetically suppressed.

\section{Discussion}
\label{Sec Discussion}

In the course of  the cooling the solitonic lines (``rods'') are growing in two stages. The first stage lasts until the rods reach some maximal characteristic size $l=l^* \sim \exp(\frac{4}{\pi} \sqrt{A_{\perp}/A_{||}})$ when the bisolitonic rods become energetically favorable with respect to the solitonic ones (or until the rods' lengths become comparable to the distance between them, which happens at $l \sim 1/\nu$). At this first stage of the growth the energy gain from merging of two rods is $\sim A\ln l$, so the growth process is self-accelerating and for finite samples the solitonic rods  can even grow across the whole sample. For big enough samples with the transverse size $H>l^*$, the single-soliton rods become eventually attracted to each other and glue into bisolitonic rods. Then the system enters the second stage of the growth, when bisolitonic rods continue to gradually elongate, but the energy gain from merging of two bisolitonic rods is only $\sim A$. So at the second stage the rods grow much more ``slowly'' with lowering the temperature. The bisolitonic rods grow to infinity as $T\rightarrow0$, so the second stage is similar to the previously studied case of the pure Ising model
\cite{Bohr:1983,Teber:2001,Karpov:2016}.
For finite systems there should be some transverse-size-dependent temperature $T_F(H)$, at which either solitonic or bisolitonic walls  grow across the whole sample.

An important role is played here by the combined amplitude-phase topological defects. The interchain order broken by an amplitude soliton $s\rightarrow -s$ (with a small core size) can be cured over the long distances by the phase degree of freedom $\theta\rightarrow\theta+\pi$. This allows the existence
of single solitons at any non-zero temperature (without confinement to bisolitonic pairs), hence there is no Ising order: $\langle s \rangle = 0$.


\section{Conclusions}
\label{Sec Conclusions}

The motivation of presented studies was to notice that most common types of symmetry breaking (superconductivity, spin density waves, antiferromagnetic Mott state, incommensurate charge density waves) in quasi-one-dimensional electronic systems possess a combined manifold of degenerate states. Beyond the standard continuous XY-type degeneracy with respect to the phase $θ$ degree of freedom (the Goldstone mode), there is also an Ising type discrete degeneracy with respect to the sign of the amplitude $A$ of the order parameter $A\exp(i\theta)$. These degrees of freedom can be controlled or accessed independently via either the spin polarization or via the charge or current densities.
The degeneracies give rise to two coexisting types of topologically nontrivial configurations: phase vortices or amplitude kinks -- the solitons. Being decoupled at the 1D level of isolated chains, the two degrees of freedom experience confinement which may be enforced or screened depending on the concentration of  solitons (given by charge or spin decompensation) and on the temperature. In 2D, 3D states with long-range (or quasi-long-range BKT type) orders, the topological confinement takes place which binds together kinks and half-integer vortices. These combined amplitude-phase solitons are the lowest-energy excitations of either spin or charge taking that role from conventional electrons.

With lowering the temperature at a finite concentration of solitons (controlled by the spin polarization or the charge doping) the combined solitonic complexes experience further aggregation resulting in a pattern formation in the course of passing through temperatures of specific phase transitions or crossovers. To understand these transformations, we have given an analytical treatment of two-fields system supported by Monte Carlo simulations for a generic coarse-grained model.
In some similarity \cite{Bohr:1983,Teber:2001,Karpov:2016} with systems possessing only the discrete degree of freedom (CDW of a bond or site dimerization type, charge ordering) here there are also two phase transitions $T_1$, $T_2$. At the higher $T_1$ the (quasi) long-range order of the XY type sets in leading to the confinement which dresses the kinks by half-integer vortices. At the lower $T_2$ the liquid of combined solitons starts to structure by aggregating the solitons into walls growing across the sample. In a 2D system, $T_2$ is only a crossover temperature below which the walls' lengths grow; but for finite samples the modeling yields sequential events of walls crossing the whole sample separating it into order parameter alternating domains. Depending on the sample width and the degree of the interchain coupling, that proceeds via either solitonic or bi-solitonic domain walls.
 The growing solitonic walls are formed as rods of transversely correlated amplitude kinks, with rods' termination points being ultimately accomplished by half-integer vortices. Attractions of vortices from different walls may force the walls to glue together forming a topologically trivial bi-solitonic wall.
Estimations and modeling results for the 3D system indicate that $T_2$ is a true phase transition.

For non-equilibrium processes, like supercooling the system with a sudden quench, the solitonic-wall scenario can be drastically promoted, because a growing solitonic rod has uncompensated half-vortex ``charges'' at the ends, which attract mobile smaller rods, while the slow process of gluing of low-mobile long solitonic rods into bisolitonic ones can be kinetically suppressed.

In view of recent successes in real space observations and even manipulations \cite{Stojchevska:2014, Yoshida:2014,Vaskivskyi:2016, Ma:2016, Cho:2016, Svetin:2016, Gerasimenko:2017, Karpov:2018} of domain walls in correlated electronic systems, we are calling for the attention to described here intriguing opportunities.

\section*{Acknowledgements}

We acknowledge the financial support of the Ministry of Education and Science of the Russian Federation in the framework of Increase Competitiveness Program of NUST MISiS (N K2-2017-085). SB acknowledges funding from the ERC AdG 320602 ``Trajectory''.

\appendix{}
\onecolumngrid

\section{Transformation from XY-picture to Coulomb gas}

Consider the simplified version of the Hamiltonian (\ref{HamiltonianIsing}) with $J_{||}=0$ and $A_{||}=A_{\perp}\equiv A$
when the Hamiltonian acquires the form
\begin{align}
H = - A \sum_r \cos(\theta_{r}-\theta_{r+\hat{x}}) - A \sum_{r} s_{r} s_{r+\hat{y}} \cos(\theta_{r}-\theta_{r+\hat{y}})
\label{Hamiltonian}
\end{align}

The partition function of the system is ($K=\beta A$):
\begin{align}
Z = \prod_{r} \int_{-\pi}^{\pi} \frac{d\theta_r}{2\pi} \exp\left( K \cos(\theta_r-\theta_{r+\hat{x}}) +
K s_r s_{r+\hat{y}} \cos(\theta_r-\theta_{r+\hat{y}}) \right)
\end{align}

We introduce ``solitonic'' bond variables $\sigma$ showing whether there is a flip of spins along a given bond in $y$-direction ($\sigma_{r,r+\hat{y}}=1$ if there are different spins at sites $r$ and $r+\hat{y}$ and $\sigma_{r,r+\hat{y}}=0$ if they are identical), and which are identically zero for $x$-direction
\begin{align}
\sigma_{r,r+\hat{y}} \equiv \sigma_{ry} =  \frac{1}{2}(1-s_r s_{r+\hat{y}}),
\sigma_{r,r+\hat{x}} \equiv \sigma_{rx} \equiv 0.
\label{sigma}
\end{align}
Denote also $\prod_{r} \int_{-\pi}^{\pi} d\theta_r/2\pi = \int d[\theta]$, and the index $\hat{\epsilon}=\hat{x},\hat{y}$, then
\begin{align}
Z = \prod_{r,\epsilon} \int d[\theta] \exp\left(K \cos(\theta_r-\theta_{r+\hat{\epsilon}} + \pi\sigma_{r\epsilon}) \right)
\end{align}

We shall employ the Villain approximation \cite{Villain:1975}: $\exp(K\cos\theta) \approx \sum_n \exp(-K(\theta-2\pi n)^2/2)$
and spread the integration over $\theta$ from ${-\infty}$ to ${\infty}$.
In this approximation spin-wave and vortical degrees of freedom are completely decoupled.
\begin{align}
Z = \sum_{\{n\}} \sum_{\{\sigma\}} \prod_{r,\epsilon} \int d[\theta] \exp\left(-\frac{K}{2} (\theta_r-\theta_{r+\hat{\epsilon}}-2\pi n_{r\epsilon}-\pi \sigma_{r\epsilon})^2 \right)
\end{align}

Now we make dual transformations, adapting the general scheme outlined in \cite{Jose:1977,Chaikin:book,Lee:1985}:

{\bf 1}. Make the Fourier transform for variables $\theta$ which gives us discrete gaussian variables $h$.

Using the identity
\begin{align}
\sum_{n=-\infty}^{\infty} \exp\left(-\frac{K}{2} (\theta-\alpha-2\pi n)^2\right)
= \frac{1}{\sqrt{2\pi K}}\sum_{h=-\infty}^{\infty} \exp\left(-\frac{h^2}{2K}\right) \exp(i h (\theta-\alpha))
\end{align}

we arrive at
\begin{align}
Z = \sum_{\{S\}} \sum_{\{\sigma\}} \prod_{r,\epsilon} \int d[\theta] \frac{1}{\sqrt{2\pi K}}
\exp\left( -\frac{h_{r\epsilon}^2}{2K}\right) \exp(i h_{r\epsilon} (\theta_r-\theta_{r+\hat{\epsilon}}-\pi \sigma_{r\epsilon}) )
\end{align}

{\bf 2}. Next we integrate over $\theta$
which yield the factor $\prod_{r} \delta(h_{rx}-h_{r-\hat{x},x} + h_{ry}-h_{r-\hat{y},y})$

\begin{figure}[!htb]
\begin{tikzpicture}
\pgfmathsetmacro{\A}{3}

\draw [ultra thick]   (0,0)  circle [radius=0.05];
\draw [ultra thick] ( \A,0)  circle [radius=0.05];
\draw [ultra thick] (-\A,0)  circle [radius=0.05];
\draw [ultra thick]  (0,\A)  circle [radius=0.05];
\draw [ultra thick]  (0,-\A) circle [radius=0.05];

\draw [dashed]  (-\A/2,-\A/2) --(-\A/2,\A/2) -- (\A/2,\A/2) -- (\A/2,-\A/2) -- (-\A/2,-\A/2);

\node [below right] at (0,0) {$r$};
\node [below,fill=white] at (-\A/2,0) {$h_{r-\hat{x},x}$};
\node [below,fill=white] at (\A/2,0) {$h_{rx}$};
\node [above right] at (0,-\A/2) {$h_{r-\hat{y},y}$};
\node [below right] at (0,\A/2) {$h_{ry}$};

\node [above left]  at (-\A/2,\A/2)  {$h_{R-\hat{x}}$};
\node [above right] at ( \A/2,\A/2)  {$h_{R}$};
\node [below right] at (\A/2,-\A/2)  {$h_{R-\hat{y}}$};
\node [below left]  at (-\A/2,-\A/2) {$h_{R-\hat{x}-\hat{y}}$};

\draw[thick] (-\A,0) -- (\A,0);
\draw[thick] (0,-\A) -- (0,\A);
\end{tikzpicture}
\\ SUPPL. FIG. 1. Original ($r$) and dual ($R$) lattices
\label{figDualLattice}
\end{figure}

{\bf 3}. Introduce integer  variables $h_R$ on the dual lattice in order to automatically satisfy the above constraint (Fig. \ref{figDualLattice}):
\begin{align*}
h_{ry} &=  h_{R-\hat{x}}-h_{R} \\
h_{rx} &= h_{R}-h_{R-\hat{y}} \\
\end{align*}
Denoting $\sigma_{rx} \equiv \sigma_{Ry} \equiv 0$, $\sigma_{ry} \equiv \sigma_{Rx}$, this yields to
\begin{align}
Z = \sum_{\{h_R\}} \sum_{\{\sigma\}} \prod_{R,\epsilon} \frac{1}{\sqrt{2\pi K}}
\exp\left( -\frac{(h_R-h_{R-\hat{\epsilon}})^2}{2K} \right) \exp(i \pi (h_R-h_{R-\hat{\epsilon}}) \sigma_{R\epsilon} )
\end{align}
If $\sigma\equiv 0$ we get the so-called ``discrete gaussian model'', which is a particular case of SOS model (with a quadratic potential). Here the terms with $\sigma\neq 0$ contribute to the partition function with ``negative weights'' -- a similar approach was explicitly used in \cite{Korshunov:1986}.
\\

{\bf 4}. Using the Poisson summation formula\cite{Kadanoff:book}
\begin{align}
\sum_{h=-\infty}^{+\infty} f(h) = \sum_{m=-\infty}^{+\infty} \int_{\phi=-\infty}^{+\infty} d\phi f(\phi) \exp(2\pi i m\phi)
\end{align}
we trade the $h$-summation to $m$-summation and $\phi$-integration; then rescaling the field $\phi_{new} = \phi/K$ we obtain
\begin{align*}
Z = \sum_{\{m_R\}} \sum_{\{\sigma\}} \prod_{R,\epsilon} \sqrt{2\pi K} \int d[\phi]
\exp\left( -\frac{K}{2} (\phi_R-\phi_{R-\hat{\epsilon}})^2 \right)
\exp(i \pi \sigma_{R\epsilon} K (\phi_R-\phi_{R-\hat{\epsilon}})) \exp(2\pi i K m_R\phi_R)
\end{align*}

{\bf 5}. Finally we can integrate out Gaussian variables $\phi$ and arrive at the Coulomb gas representation.
\begin{align*}
\int D\phi \exp\left(-\frac{1}{2} \sum_{i,j} \phi_i A_{ij} \phi_j - \sum_i J_i \phi_i\right) \sim
(\det A)^{-1/2} \exp\left(\frac{1}{2} \sum_{i,j} J_i (A^{-1})_{ij} J_j\right)
\end{align*}
\begin{align*}
K \sum_{\langle R,R' \rangle} (\phi_R-\phi_{R'})^2 = \sum_{R,R'} \phi_R \underbrace{K (4 \delta_{R,R'}-\delta_{R,R'+\hat{x}}-\delta_{R,R'-\hat{x}} -\delta_{R,R'+\hat{y}}-\delta_{R,R'-\hat{y}})}_{A_{RR'}} \phi_{R'}
\end{align*}
\begin{align*}
A_{\mathbf k} = \sum_R e^{-i {\mathbf k} {\mathbf R}} A_{R,0} = K(4 - 2\cos(k_x) -2\cos(k_y))
\end{align*}
\begin{align*}
(A^{-1})_{{\mathbf R}{\mathbf R}'} = \int d^2 k e^{i{\mathbf k} ({\mathbf R}-{\mathbf R}')} (A^{-1})_{\mathbf k} =  \frac{1}{K}\int d^2 k \frac{e^{i{\mathbf k} ({\mathbf R}-{\mathbf R}')}}{4 - 2\cos(k_x) -2\cos(k_y)} \sim -\frac{1}{2\pi K} \ln \frac{|{\mathbf R}-{\mathbf R}'|}{a} + \frac{1}{4K},
\end{align*}
where $a$ is the cutoff of the order of the lattice spacing.
Note that the Ising variables $\sigma$ do not appear in the matrix $A$; they enter only the ``current'' term
\begin{align*}
i J_R \phi_R \equiv 2\pi i K (m_R+\nu_R) \phi_R
\end{align*}
Here we denoted half-integer contributions to vorticity as $\nu_R = (\sigma_{R,x}-\sigma_{R+\hat{x},x} +\sigma_{R,y}-\sigma_{R+\hat{y},y})/2$. We see that the standard XY-vorticity is just shifted by $\nu_R = 0, \pm 1/2$, which
defines
the Coulomb gas with a mixture of integer and half-integer charges. Completing the Gaussian integration over $\phi$ and substituting back $K=\beta A$ we obtain
\begin{align}
Z = \sum_{\{m_R\}} \sum_{\{\nu(s)\}}
\exp\left( \pi \beta A \sum_{R\neq R'} (m_R + \nu_R) \ln\frac{|{\mathbf R}-{\mathbf R}'|}{a} (m_{R'} + \nu_{R'}) + (\ln y_0) \sum_R (m_R+\nu_R)^2  \right)
\label{Z_Coulomb}
\end{align}
Here  $y_0 = \exp(-\pi^2 \beta A/2)$, the sum $\sum_{\{m_R\}}$ contains only neutral configurations $\sum_R (m_R+\nu_R) = 0$, the sum $\sum_{\{\nu(s)\}}$ counts all possible configurations of spins $\{s\}$ with corresponding values of $\nu(\{s\})$ (recall the definition of $\sigma_{ry}$ (\ref{sigma}))
\begin{align}
\nu_R = (\sigma_{Rx}-\sigma_{R+\hat{x},x})/2 = (\sigma_{ry}-\sigma_{r+\hat{x},y})/2 = \frac{1}{4}(s_{r+\hat{x}} s_{r+\hat{x}+\hat{y}} - s_r s_{r+\hat{y}})
\label{nu_R}
\end{align}

So, we have constructed a dual model  with two types of vorticity variables $m_R$ and $\nu_R$, making the vorticities effectively half-integer. We use the partition function (\ref{Z_Coulomb}) in Sec. \ref{sec_analytical} for constructing the strong coupling part of the phase diagram corresponding to $J_{||}=0$.

\section{Modeling for 2D system $100\times100$ with strong coupling $A_{\perp} = 4 A_{||}$}

\begin{figure}[H]
	\centering
	\includegraphics[width=0.7\linewidth]{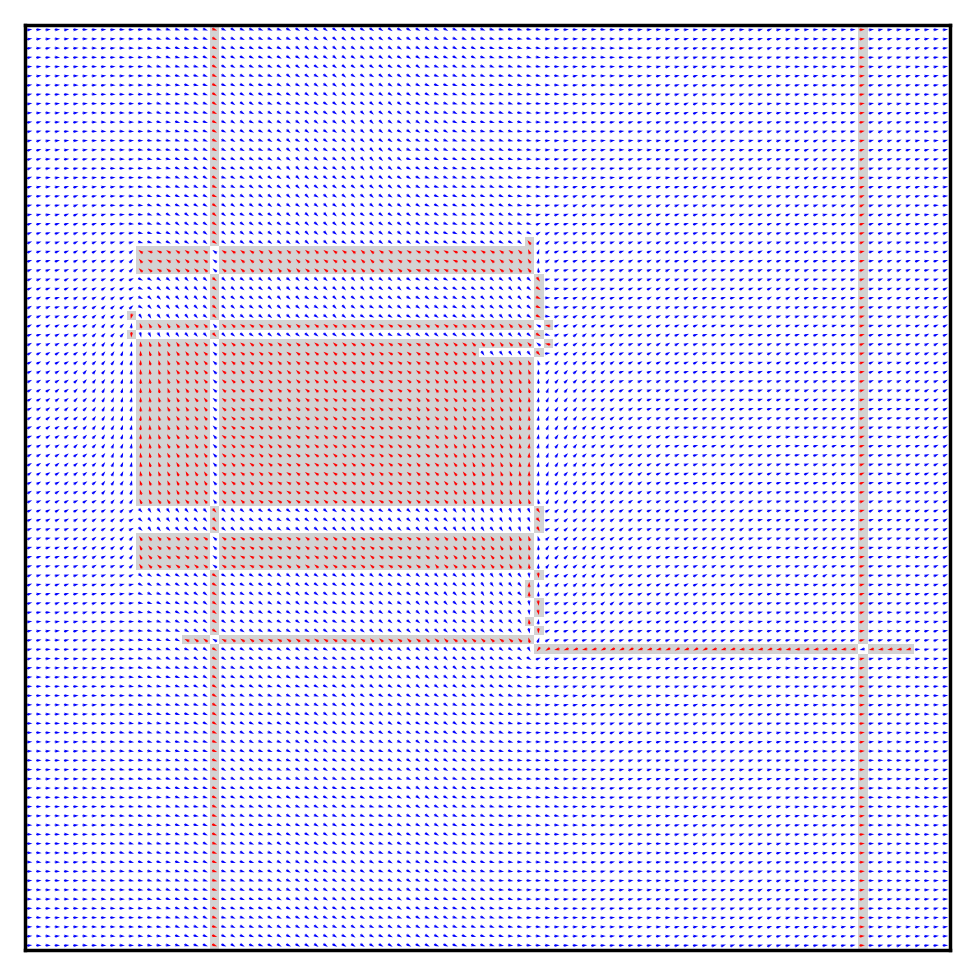}%
\\	SUPPL. FIG. 2. Configuration of the system $100\times100$ with $A_{\perp} = 4 A_{||}$ at $T=0.01 A$.
\label{fig_Appendix}
\end{figure}

\twocolumngrid


\begin{thebibliography}{99}
\bibitem{Landau-argument}
For arbitrary small $T\neq0$ it is energetically favorable to break the in-chain order by such finite-energy defects, since the entropy term $S\sim \ln L$ overcomes the energy term and lowers the free energy $F=E-TS$ as $L\rightarrow\infty$.
\end{thebibliography}
\end{document}